\documentclass[]{aastex631}

\usepackage{url}
\usepackage{amsmath}
\usepackage{graphicx}
\usepackage{booktabs}
\usepackage{natbib}

\graphicspath{{./}{figures/}} 

\def\figref#1{Figure~\ref{#1}} 
\def\tabref#1{Table~\ref{#1}} 
\def\eqref#1{Equation~\ref{#1}} 
\def\secref#1{Section~\ref{#1}} 

\def\mycheck#1{\textcolor{red}{#1???}} %
\def\mycomment#1{\textcolor{cyan}{\bf [NT] #1}} %
\def\revise#1{\textcolor{red}{\bf #1}}

\def\lhaasoj{LHAASO~J0341$+$5258}

\def\xmm{{\it XMM-Newton}}

\def\rosat{{\it ROSAT}}

\def\lat{{\it Fermi}-LAT}

\newcommand{\nobeyama}{the NRO 45-m telescope}

\def\dag{\hbox{$^\dagger$}}
\def\ddag{\hbox{$^\ddagger$}}

\def\eflux{\hbox{${\rm erg}~{\rm cm}^{-2}~{\rm s}^{-1}$}} 

\def\kms{\hbox{${\rm km}~ {\rm s}^{-1}$}}
\def\cc{\hbox{${\rm cm}^{-3}$}}
\def\columnd{\hbox{${\rm cm}^{-2}$}}  

\newcommand\HI{HI}

\newcommand\COa{$^{12}$CO($J$=1--0)}
\newcommand\COb{$^{13}$CO($J$=1--0)}
\newcommand\COc{C$^{18}$O($J$=1--0)}

\newcommand\Vlsr{V$_{\rm LSR}$}
\newcommand\TMB{$T_{\rm MB}$}

\def\NHtwo{\hbox{$N({\rm H_2}$)}}

\def\sun{\hbox{$\odot$}}
\def\Msun{\hbox{$M_\odot$}}

\usepackage{acro}

\DeclareAcronym{gde}{
  short = GDE ,
  long  = Galactic Diffuse Emission,
}

\DeclareAcronym{cgb}{
  short = CGB ,
  long  = cosmic gamma-ray background,
}

\DeclareAcronym{grb}{
  short = GRB ,
  long  = gamma-ray burst,
}
 
\DeclareAcronym{agn}{
  short = AGN ,
  long  = active galactic nucleus ,
  long-plural-form = active galactic nuclei
}
\DeclareAcronym{fsrq}{
  short = FSRQ ,
  long  = flat-spectrum radio quasar,
}
  
\DeclareAcronym{lmxb}{
  short = LMXB ,
  long  = low mass X-ray binary ,
  long-plural-form = low mass X-ray binaries ,
}
\DeclareAcronym{hmxb}{
  short = HMXB ,
  long  = high mass X-ray binary ,
  long-plural-form = high mass X-ray binaries ,
}
  
\DeclareAcronym{ic}{
  short = IC ,
  long  = inverse Compton ,
}

\DeclareAcronym{nustar}{
  short = {\it NuSTAR} ,
  long  = Nuclear Spectroscopic Telescope Array ,
}

\DeclareAcronym{bi}{
  short = BI ,
  long  = backside illumination ,
}
\DeclareAcronym{fi}{
  short = FI ,
  long  = frontside illumination ,
}
\DeclareAcronym{fov}{
  short = FoV ,
  long  = field of view ,
}
  
\DeclareAcronym{sim}{
  short = SIM ,
  long  = Science Instrument Module ,
}
\DeclareAcronym{hetg}{
  short = HETG ,
  long  = High Energy Transmission Grating ,
}
\DeclareAcronym{letg}{
  short = LETG ,
  long  = Low Energy Transmission Grating ,
}
\DeclareAcronym{hrc}{
  short = HRC ,
  long  = High Resolution Camera ,
}
\DeclareAcronym{acis}{
  short = ACIS ,
  long  = Advanced CCD Imaging Spectrometer ,
}
\DeclareAcronym{hrma}{
  short = HRMA ,
  long  = High Resolution Mirror Assembly,
}

\DeclareAcronym{compton}{
  short = Compton ,
  long  = Compton Gamma Ray Observatory,
}
\DeclareAcronym{hst}{
  short = HST ,
  long  = Hubble Space Telescope ,
}
\DeclareAcronym{iact}{
  short = IACT ,
  long  = Imaging Atmospheric Cherenkov Telescope ,
}
\DeclareAcronym{wcd}{
  short = WCD ,
  long  = Water Cherenkov Detector ,
}

\DeclareAcronym{hawc}{
  short = HAWC ,
  long  = High-Altitude Water Cherenkov ,
}
\DeclareAcronym{cta}{
  short = CTA ,
  long  = Cherenkov Telescope Array ,
}

\DeclareAcronym{em}{
  short = EM ,
  long  = electromagnetic ,
}
\DeclareAcronym{ism}{
  short = ISM ,
  long  = interstellar medium ,
}
\DeclareAcronym{csm}{
  short = CSM ,
  long  = circumstellar medium ,
}
\DeclareAcronym{sne}{
  short = SNe ,
  long  = supernovae , 
}

\DeclareAcronym{iss}{
  short = ISS ,
  long  = International Space Station ,
}

\DeclareAcronym{uhecr}{
  short = UHECRs ,
  long  = ultra high energy cosmic rays , 
}
\DeclareAcronym{ta}{
  short = TA ,
  long  = Telescope Array , 
}
\DeclareAcronym{auger}{
  short = Auger ,
  long  = Pierre Auger Observatory , 
}
\DeclareAcronym{ams}{
  short = AMS ,
  long  = Alpha Magnetic Spectrometer , 
}
\DeclareAcronym{pamela}{
  short = PAMELA ,
  long  = Payload for Antimatter Matter Exploration and Light-nuclei Astrophysics , 
}
   
\DeclareAcronym{cmb}{
  short = CMB ,
  long  = Cosmic Microwave Background , 
}
\DeclareAcronym{sed}{
  short = SED ,
  long  = spectral energy distribution , 
}
\DeclareAcronym{mhd}{
  short = MHD ,
  long  = magnetohydrodynamical ,
}
\DeclareAcronym{dof}{
  short = dof ,
  long  = degree of freedom ,
}
\DeclareAcronym{cco}{
  short = CCO ,
  long  = central compact object ,
  first-style = default
}
\DeclareAcronym{lmc}{
  short = LMC ,
  long  = Large Magellanic Cloud ,
}
\DeclareAcronym{smc}{
  short = SMC ,
  long  = Small Magellanic Cloud ,
}

\DeclareAcronym{hess}{
  short = H.E.S.S. ,
  long  = High Energy Stereoscopic System ,
  first-style = default
}
\DeclareAcronym{snr}{
  short = SNR ,
  long  = supernova remnant ,
}
\DeclareAcronym{pwn}{
  short = PWN ,
  short-plural = e ,
  long  = pulsar wind nebula ,
  long-plural  = e ,
}

\DeclareAcronym{sn}{
  short = SN ,
  short-plural = e ,
  long  = supernova ,
  long-plural  = e ,
  first-style = default
}

\DeclareAcronym{nw}{
  short = NW ,
  long  = northwest ,
  first-style = default
}
\DeclareAcronym{hxc}{
  short = HXC ,
  long  = hard X-ray component ,
  first-style = default
}

\DeclareAcronym{cr}{
  short = CR ,
  long  = cosmic ray ,
}
\DeclareAcronym{psf}{
  short = PSF ,
  long  = point spread function ,
}
\DeclareAcronym{hpd}{
  short = HPD ,
  long  = half power diameter ,
}
\DeclareAcronym{fwhm}{
  short = FWHM ,
  long  = full width of half maximum ,
}

\DeclareAcronym{pic}{
  short = PIC ,
  long  = particle-in-cell ,
  tag = numerical ,
}
\DeclareAcronym{cxb}{
  short = CXB ,
  long  = Cosmic X-ray Background ,
}
\DeclareAcronym{grxe}{
  short = GRXE ,
  long  = Galactic Ridge X-ray Emission ,
}
\DeclareAcronym{pa}{
  short = PA ,
  long  = Positional Angle ,
}
\DeclareAcronym{dsa}{
  short = DSA ,
  long  = diffusive shock acceleration ,
}

\def\degr{\hbox{$^\circ$}}
\def\arcmin{\hbox{$^\prime$}}
\def\arcsec{\hbox{$^{\prime\prime}$}}
\def\utw{\smash{\rlap{\lower5pt\hbox{$\sim$}}}}
\def\udtw{\smash{\rlap{\lower6pt\hbox{$\approx$}}}}
\def\apjl{ApJ}%
\def\apjs{ApJS}%
\def\ao{Appl.~Opt.}%
\def\aap{A\&A}%
\acsetup{patch / longtable = false } 

\begin{document}

\title{
Detection of molecular clouds in the PeVatron candidate source LHAASO J0341$+$5258 by the Nobeyama 45-m radio telescope
}

\author[0000-0001-7209-9204]{Naomi Tsuji}
\affiliation{Faculty of Science, Kanagawa University, 3-27-1 Rokukakubashi, Kanagawa-ku, Yokohama-shi, Kanagawa 221-8686}
\affiliation{Interdisciplinary Theoretical \& Mathematical Science Program (iTHEMS), RIKEN, 2-1, Hirosawa, Saitama 351-0198, Japan}
\affiliation{Department of Physics, Rikkyo University, Nishi-Ikebukuro 3-34-1, Toshima-ku, Tokyo, 171-8501, Japan}
\author[0000-0001-8147-6817]{Shunya Takekawa}
\affiliation{Department of Applied Physics, Faculty of Engineering, Kanagawa University, 3-27-1 Rokkakubashi, Kanagawa-ku, Yokohama, Kanagawa 221-8686, Japan}
\author[0000-0002-9709-5389]{Kaya Mori} 
\affiliation{Columbia Astrophysics Laboratory, Columbia University, 538 West 120th Street, New York, NY 10027, USA}
\author[0000-0003-3631-5648]{Alison Mitchell} \affiliation{Erlangen Centre for Astroparticle Physics, Friedrich-Alexander-Universit\"{a}t Erlangen-N\"{u}rnberg, Nikolaus-Fiebiger-Str. 2, Erlangen 91058, Germany}
\author[0000-0002-2967-790X]{Shuo Zhang} \affiliation{Department of Physics and Astronomy, Michigan State University, East Lansing, MI, 48824, USA} 
\author[0000-0002-3886-3739]{Priyadarshini Bangale},\affiliation{Department of physics, Temple University, Philadelphia, PA 19122, USA}
\author[0000-0003-2633-2196]{Stephen DiKerby} \affiliation{Department of Physics and Astronomy, Michigan State University, East Lansing, MI, 48824, USA} 
\author[0000-0003-2423-4656]{T\"ul\"un Ergin} \affiliation{Department of Physics and Astronomy, Michigan State University, East Lansing, MI, 48824, USA} 
\author[0009-0001-6471-1405]{Jooyun Woo}  \affiliation{Columbia Astrophysics Laboratory, Columbia University, 538 West 120th Street, New York, NY 10027, USA}
\author[0000-0001-6189-7665]{Samar Safi-Harb}
\affiliation{Department of Physics and Astronomy, University of Manitoba, Winnipeg, MB R3T 2N2, Canada}
\author[0009-0004-4904-5792]{Shinta Kasuya} \affiliation{Faculty of Science, Kanagawa University, 3-27-1 Rokukakubashi, Kanagawa-ku, Yokohama-shi, Kanagawa 221-8686}
%



\begin{abstract}

We report a new CO observation survey of LHAASO J0341$+$5258 using the Nobeyama Radio Observatory (NRO) 45-m telescope. 
LHAASO J0341$+$5258 is one of the unidentified ultra-high-energy (UHE; E $>$100 TeV) gamma-ray sources detected by LHAASO. 
Our CO observations were conducted in February and March 2024, with a total observation time of 36 hours, covering the LHAASO source ($\sim$0.3--0.5 degrees in radius) and its surrounding area (1$\times$1.5 degrees). 
Within the LHAASO source extent, we identified five compact ($<$ 2 pc) molecular clouds at nearby distances ($<$ 1--4 kpc). 
These clouds can serve as proton-proton collision targets, producing hadronic gamma rays via neutral pion decays. 
Based on the hydrogen densities (700--5000 cm$^{-3}$) estimated from our CO observations and archived HI data from the DRAO survey, we derived the total proton energy of $W_p$ (E $>$ 1 TeV) $\sim$ 10$^{45}$ erg to account for the gamma-ray flux. 
One of the molecular clouds appears to be likely associated with an asymptotic giant branch (AGB) star with an extended CO tail, which may indicate some particle acceleration activities. However, the estimated maximum particle energy below 100 TeV makes the AGB-like star unlikely to be a PeVatron site. 
We conclude that the UHE emission observed in LHAASO J0341$+$5258 could be due to hadronic interactions between the newly discovered molecular clouds and TeV-PeV protons originating from a distant SNR 
or due to leptonic emission from a pulsar wind nebula candidate, which is reported in our companion X-ray observation paper (DiKerby et al. 2025). 

\end{abstract}

\keywords{
ISM: clouds --- 
radio lines: ISM ---
ISM: individual objects (\lhaasoj) --- 
cosmic rays --- 
gamma rays 
}


\section{Introduction} \label{sec:intro}
Cosmic rays (CRs), primarily high-energy protons, with energies below the \textit{knee} of their spectrum --- approximately a few PeV --- have been proposed to originate from supernova remnants (SNRs) in our Galaxy.
The so called ``pion bump'' structure, which is a smoking gun of accelerated protons, has been likely identified in gamma-ray spectra of several SNRs \citep{Fermi2013_pi0}. 
This discovery has led to SNRs being widely accepted as primary contributors to CR protons.
However, it is lacking decisive observational evidence whether SNRs are indeed capable of accelerating particles up to the knee, despite numerous dedicated studies \citep[e.g., ][]{Tanaka2008,HESS2018_SNR,abeysekara_evidence_2020,tsuji_systematic_2021}.
In fact, the gamma-ray spectra of most SNRs have exponential cutoffs at TeV \citep{Funk2015}, making it challenging to reconcile with particle acceleration up to PeV. 
One possible resolution is that SNRs act as PeVatrons only for a short period early in their evolution.

Recently, (sub-)PeV gamma-ray astronomy has begun, driven by the detection of ultra high-energy (UHE; $E>$100 TeV) gamma rays by air shower arrays such as Tibet AS$\gamma$, HAWC, and LHAASO \citep{amenomori_northern_2005,hawc_collaboration_multiple_2020,lhaaso_collaboration_future_2010}.
The findings by these observatories may be on the verge of changing our understanding of the SNR paradigm (SNRs as primary CR accelerators), especially for CRs with PeV energies.
Measurements of UHE gamma rays can help us pinpoint the location of CRs with PeV energies and plausible accelerators of PeV CRs (i.e., PeVatrons) in the vicinity.
More recently, stellar clusters and microquasars have emerged as new classes of PeVatrons after some of them were detected in the UHE gamma-ray band \citep[e.g., ][]{cao_ultrahigh-energy_2021,aharonian_deep_2022,hess_collaboration_acceleration_2024,lhaaso_2024_cygnus,alfaro_ultra-high-energy_2024,lhaaso_2024_microquasars}.
On the other hand, the UHE emission in the W51 complex likely provides the first evidence of particle acceleration up to the PeV energy range in SNR W51C \citep{cao_evidence_2024}.

Nevertheless, there is still a non-negligible number of unidentified TeV-PeV gamma-ray sources.
For example, about 40\% of the 90 sources listed in the first LHAASO catalog \citep{cao_first_2024} have no apparent associations and remain  unidentified.
Some of them remain unidentified simply due to a lack of observational data at other wavelengths.
Radio and X-ray follow-up observations already proved the importance of multiwavelength study such as in HESS J1702$-$420 \citep{giunti_constraining_2022} and LHAASO J2108$+$5157 \citep{de_la_fuente_detection_2023}.

There exists a long-standing question about the nature of gamma-ray sources; leptonic (inverse Compton scattering (ambient photons up-scattered by accelerated electrons) or nonthermal bremsstrahlung) or hadronic (decay of pions produced in interactions between accelerated protons and low-energy protons in ambient gas).
Identifying gamma rays as hadronic emission is meaningful in the context of CRs, which predominantly consist of energetic protons.
Hadronic gamma rays should be accompanied by dense gas, such as atomic and molecular clouds, which can be revealed by radio observations.
Detecting neutrinos produced in the hadronic process would be another evidence of proton acceleration.
Furthermore, in the hadronic scenario, the same proton-proton interactions produce charged pions, decaying into muons and secondary electrons and positrons.
Thus, measuring the synchrotron radiation from these secondary electrons can be a strong indication for hadronic gamma rays, although it has not been detected yet \citep{aharonian_gamma_2013,tsuji_search_2024}.

\begin{figure}[ht!]
 \begin{center}
  \includegraphics[width=0.7\linewidth]{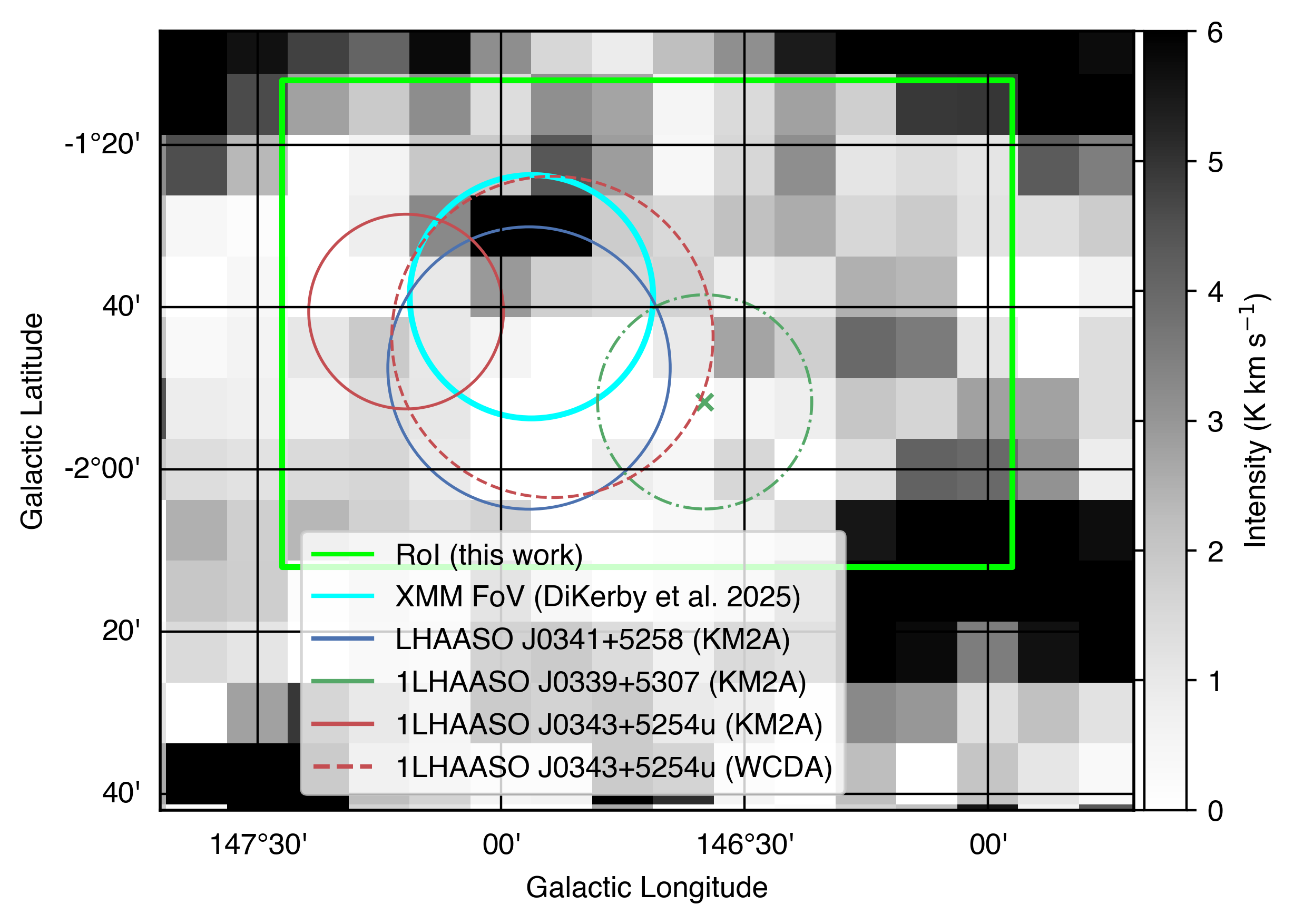} 
 \end{center}
\caption{
The CfA \COa\ map \citep{dame_milky_2001,dame_co_2022}, integrated from $-$40 \kms\ to 10 \kms, in the \lhaasoj\ region.
The blue and red circles show the gamma-ray extent of LHAASO J0341$+$5258 and 1LHAASO J0343$+$5254u, respectively.
For 1LHAASO J0339$+$5307, which was detected as a point-like source, the best-fit position and the 95\% statistical upper limit of its radius are shown with the dark green cross and the dash-dotted circle, respectively.
The green rectangle indicates the region of 1.5$\times$1 degree$^2$ which we scanned with \nobeyama.
The cyan circle shows the field of view of the \xmm\ data \citep{dikerby2024}.
}
\label{fig:overview}
\end{figure}

To advance our understanding of the PeVatron population, we need to investigate the unidentified UHE gamma-ray sources and reveal their origins one by one.
In this paper, we focus on one of the PeVatron candidates and unidentified sources, LHAASO J0341$+$5258 (hereafter J0341). 
This source was discovered by LHAASO's Kilometer Squared Array (KM2A) as an extended source, as summarized in \tabref{tab:overview} \citep{cao_discovery_2021}.
In the following catalog \citep{cao_first_2024}, it was split into two sources, 1LHAASO J0339$+$5307 (only detected by KM2A) and 1LHAASO J0343$+$5254u (detected by both KM2A and Water Cherenkov Detector Array (WCDA)) (see \tabref{tab:overview} and \figref{fig:overview} for details).
Note that 1LHAASO J0343$+$5254u is detected in the UHE gamma-ray band. 
Although this region seems to be crowded with two or three LHAASO source components, the overall spectrum is hard ($\Gamma\approx 1.7$) in the 1--25 TeV energy range and becomes softened ($\Gamma \approx 3.5$) above 25 TeV, indicating the existence of a cutoff or curvature. Indeed, the spectrum in \cite{cao_discovery_2021} appeared to be better fit by a cutoff power-law or log-parabola model.
Later, \cite{bangale_searching_2023} reported the gamma-ray detection in 10--200 TeV by HAWC and upper limit in 0.5--50 TeV by VERITAS.
Within the gamma-ray extent measured by LHAASO, there is an unidentified GeV gamma-ray emission region, 4FGL J0340.4$+$5302, showing a curved spectrum with a peak energy around 200 MeV \citep{cao_discovery_2021,4fgl}.
\cite{kar_ultrahigh-energy_2022} and \cite{de_sarkar_dissecting_2023} conducted detailed modelings, inferring that the maximum CR energy would be roughly 200 TeV either in leptonic or hadronic.
Although there are three \rosat\ X-ray point-like sources within the gamma-ray extension, there is no known extended X-ray emission so far.
As part of an approved AO-22 \xmm\ Large Program, the J0341 region was observed in February 2024 (\figref{fig:overview}), and an associated publication is in preparation \citep{dikerby2024}.

In this paper, we search for molecular clouds in the PeVatron candidate source, J0341, using the Nobeyama Radio Observatory (NRO) 45-m radio telescope.
We present an observation overview in Section~\ref{sec:observation} and the analysis and results in Section~\ref{sec:results}.
We also report the results of analysis of the archival \HI\ and radio continuum data in \secref{sec:CGPS}.
The discussion and conclusions are described in  Sections~\ref{sec:discussion} and \ref{sec:conclusions}, respectively.

\begin{table}[h!]
  \begin{center}
  \caption{LHAASO sources in the J0341 region}
  \begin{tabular}{cccccccc}
      \hline
Name              &  Instrument$^*$    & ($\ell$, $b$)$^\dagger$  & Size$^\S$   & $\Gamma$ & TS$_{>100~ \mathrm{TeV}}^\ddagger$ & Reference \\
              &      &  (deg) &   (deg) &  &  &  \\
\hline 
LHAASO J0341$+$5258  & KM2A & (146.94, $-$1.792)  & 0.29$\pm$0.06 &  2.98$\pm$0.19 & --- & \cite{cao_discovery_2021} \\
\hline
1LHAASO J0339$+$5307 & KM2A & (146.58, $-$1.861)  & $<$0.22 & 3.64$\pm$0.16 & --- &  \cite{cao_first_2024} \\
                   & WCDA & ---  & --- & --- & --- &  \cite{cao_first_2024} \\
\hline
1LHAASO J0343$+$5254u & KM2A & (147.20, $-$1.676) & 0.20$\pm$0.02 & 3.53$\pm$0.10 & 20.2 &  \cite{cao_first_2024} \\
                    & WCDA & (146.89, $-$1.728) & 0.33$\pm$0.05 & 1.70$\pm$0.19 &  --- & \cite{cao_first_2024} \\
\hline
    \end{tabular}
    \label{tab:overview}
  \end{center}
    \tablecomments{ \\
$^*$ The energy range of KM2A is $>$ 25 TeV, and that of WCDA is 1--25 TeV. \\
$^\dagger$ The uncertainty on the position is $\sim$0.1\degr. \\
$^\S$ Size indicates $r_{39}$ (39\% containment radius of the 2D-Gaussian model), except that the 95\% upper limit is provided for 1LHAASO J0339$+$5307.  \\ 
$^\ddagger$ TS$_{>100~ \mathrm{TeV}}$ is a test statistic value of detection above 100 TeV.
}
\end{table}

\section{Observations} \label{sec:observation}

The \COa, \COb, and \COc\ data were obtained with the NRO 45-m radio telescope in February and March in 2024.  
We scanned an area of 1.5 $\times$ 1 degree$^2$, centered at $(\ell, ~ b)$ = (146.7\degr, $-$1.7\degr), as shown in \figref{fig:overview}. 
The scanning was made along the Galactic latitude, and the observations were performed in on-the-fly mapping mode \citep{sawada_--fly_2008}.
The total observation time was 36 hours.
The spatial resolution of \nobeyama\ is 15\arcsec, which is much improved compared to the existing survey data (e.g., the CfA map with the angular resolution of 0.5\degr\ \citep{dame_milky_2001}).

We made use of the four-beam receiver FOREST \citep{minamidani_development_2016} and the autocorrelation spectrometer SAM45 \citep{kuno_new_2011}. 
The system noise temperatures including the atmosphere were between 300 and 400 K at 115 GHz. 
The bandwidth and resolution were 122.07 kHz both at 115 GHz and 110 GHz. 
The half power beam width (HPBW) is 14 arcsec at 115 GHz and 15 arcsec at 110 GHz\footnote{\url{https://www.nro.nao.ac.jp/~nro45mrt/html/prop/eff/eff_latest.html}\label{fot:nro}}.
The observations were made by the 2:1 
ON-OFF position switching mode.
The OFF points were centered at $(\ell, ~ b)$ = (148.5\degr, $-$1.9\degr) for $\ell \geq 146.5\degr$ and $(\ell, ~ b)$ = (143.7\degr, $-$2.9\degr) for $\ell \leq 146.5\degr $. 
The pointing accuracy was checked every $\sim$2 hours by observing the SiO maser, S-per, at the position of $(\ell, ~ b)$ = (134.62\degr, $-$2.1949\degr), using the H40 receiver.
We checked that the pointing accuracy was better than 3\arcsec\ in both azimuth and elevation.
Calibration of the antenna temperature ($T_a^*$) was accomplished by the standard chopper-wheel method \citep{kutner_recommendations_1981}.

All data were reduced using the NOSTAR reduction package.
To determine and subtract the baseline, we conducted a linear fitting in ranges from $-$200 to $-$150 \kms\ and from 150 to 200 \kms.
The antenna temperature ($T_a^*$) was converted to main-beam temperature ($T_{\rm MB}$) by dividing by a main-beam efficiency ($\eta_{\rm MB}$) as $T_{\rm MB} = T_a^* / \eta_{\rm MB}$.
The applied $\eta_{\rm MB}$, provided by the observatory\footref{fot:nro}, was 0.35 at 115 GHz and 0.40 at 110 GHz.
As an intensity calibration source, we observed a dense cloud region seen in \cite{cao_ultrahigh-energy_2021}, which is characterized by a circle centered at $(\ell, ~ b)$ = (147.05\degr, $-$1.5098\degr) and a radius of 1.25\arcmin.
The flux of this calibration source was fairly constant (less than 10\%) during the observation dates, thus we combined all the data at different epochs without scaling.

Maps were produced by convolution using Bessel-Gaussian functions.
The spatial and velocity grids have sizes of 7.5\arcsec\ and 1 \kms, respectively. 
The velocity coverage is from $-$150 to 150 \kms.
The root-mean-square (RMS) noise level in \TMB\ is 1.15 K and 0.324 K at 115 and 110 GHz, respectively. 
We applied 3-RMS cut to reduce the noise level for the following analyses.

\section{Analysis and results}\label{sec:results}

\begin{figure*}[ht!]
 \begin{center}
  \includegraphics[width=0.8\linewidth]{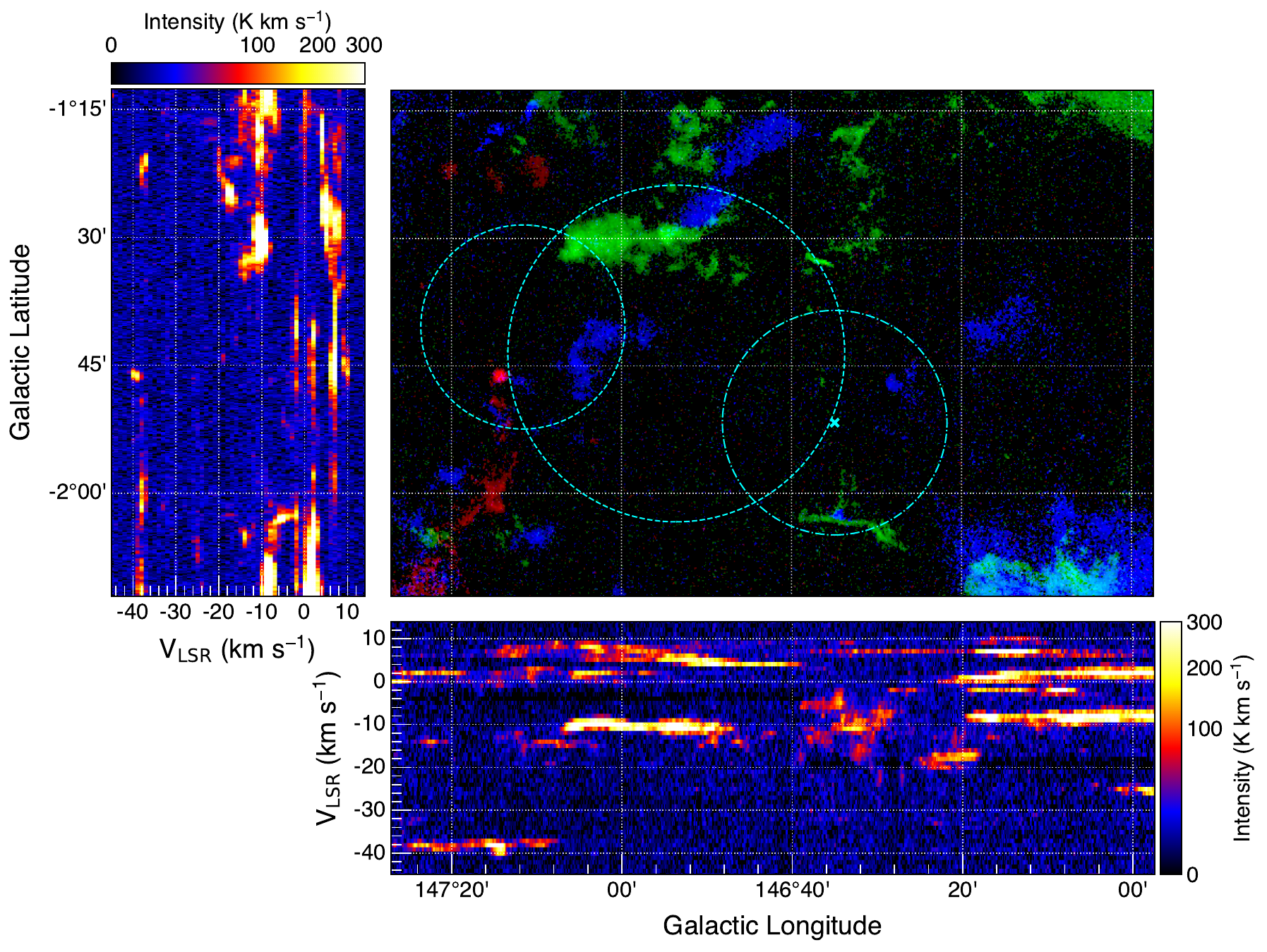}
 \end{center}
\caption{\COa\ RGB image and position-velocity maps.
In the RGB image, red, green, and blue indicate the velocity integrated maps in 
$-$41 to $-$37~\kms,
$-$15 to $-$5~\kms,
$-$3 to 5~\kms, respectively.
The cyan circles indicate the 1LHAASO sources \citep{cao_first_2024}, as summarized in \figref{fig:overview}.
In the Galactic latitude (longitude)-velocity map, the intensity is integrated over the entire region along the Galactic longitude (latitude).
}
\label{fig:map12}
\end{figure*}

\begin{figure*}[ht!]
 \begin{center}
  \includegraphics[width=0.8\linewidth]{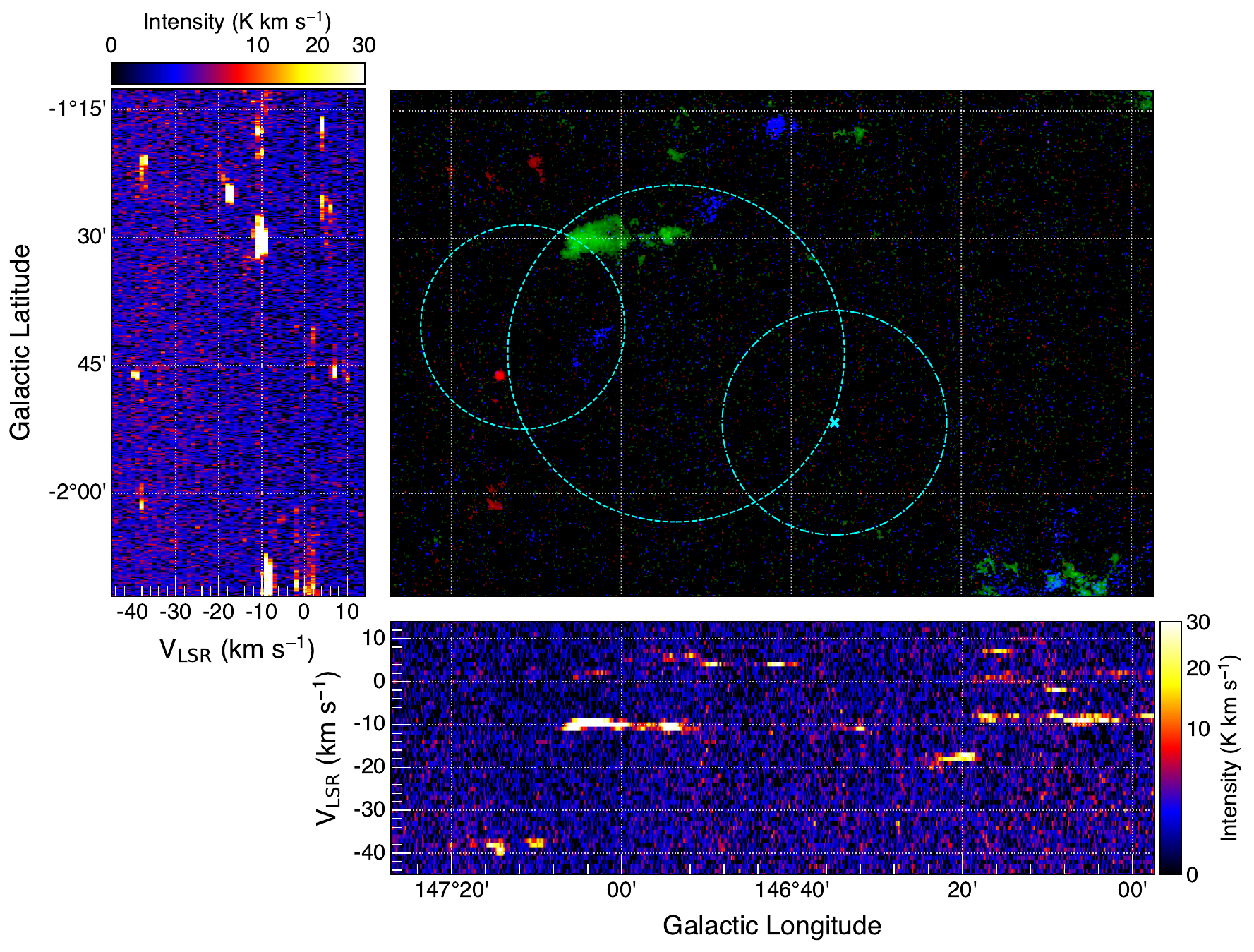}
 \end{center}
\caption{The same as \figref{fig:map12}, but for \COb.}
\label{fig:map13}
\end{figure*}

Figure~\ref{fig:map12} shows the velocity-integrated and position-velocity maps of \COa, and \figref{fig:map13} shows those of \COb, obtained by \nobeyama.
%
Although the velocity coverage of our observations is from $-$150 to 150 \kms, there is no significant signal at \Vlsr $< -40$~\kms\ and \Vlsr $> 10$~\kms.
As illustrated in Figures~\ref{fig:map12} and \ref{fig:map13} (see also Figures~\ref{fig:grid} and \ref{fig:spectrum} in Appendix~\ref{sec:app}), there exist some clouds at different velocities in the region of interest (RoI),
and we divided them into three groups: 
(1) $-$41~\kms\ $\leq$ \Vlsr\ $< ~ -$37~\kms, 
(2) $-$15~\kms\ $\leq$ \Vlsr\ $< ~ -$5~\kms, and
(3) $-$3~\kms\ $\leq$ \Vlsr\ $< ~ $5~\kms.

\begin{figure}[h!]
 \begin{center}
  \includegraphics[width=0.6\linewidth]{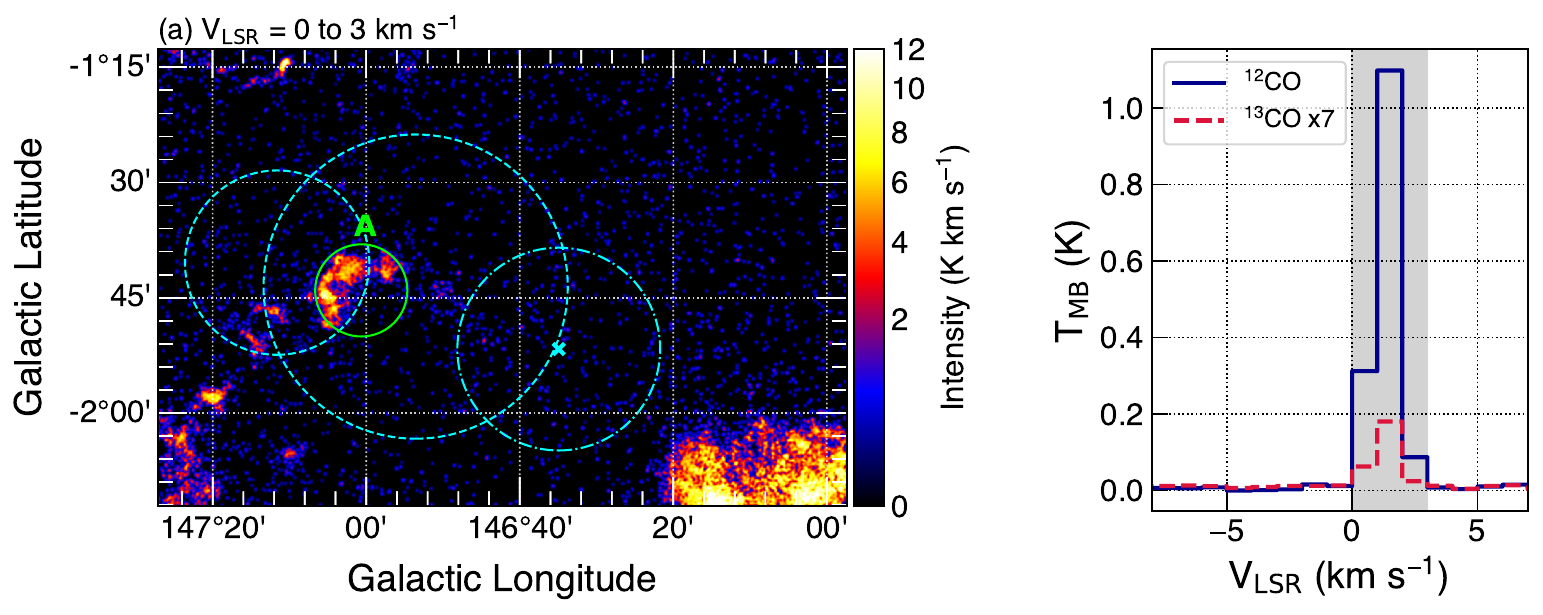}
  \includegraphics[width=0.6\linewidth]{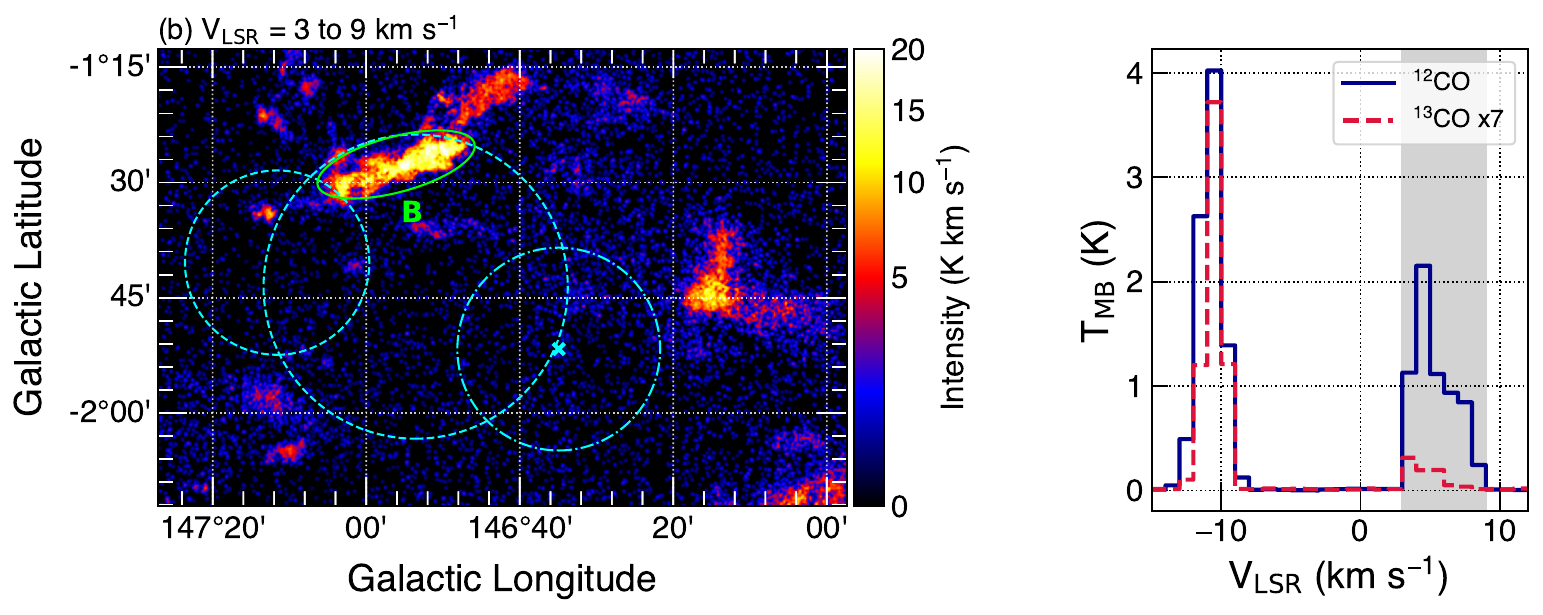}
  \includegraphics[width=0.6\linewidth]{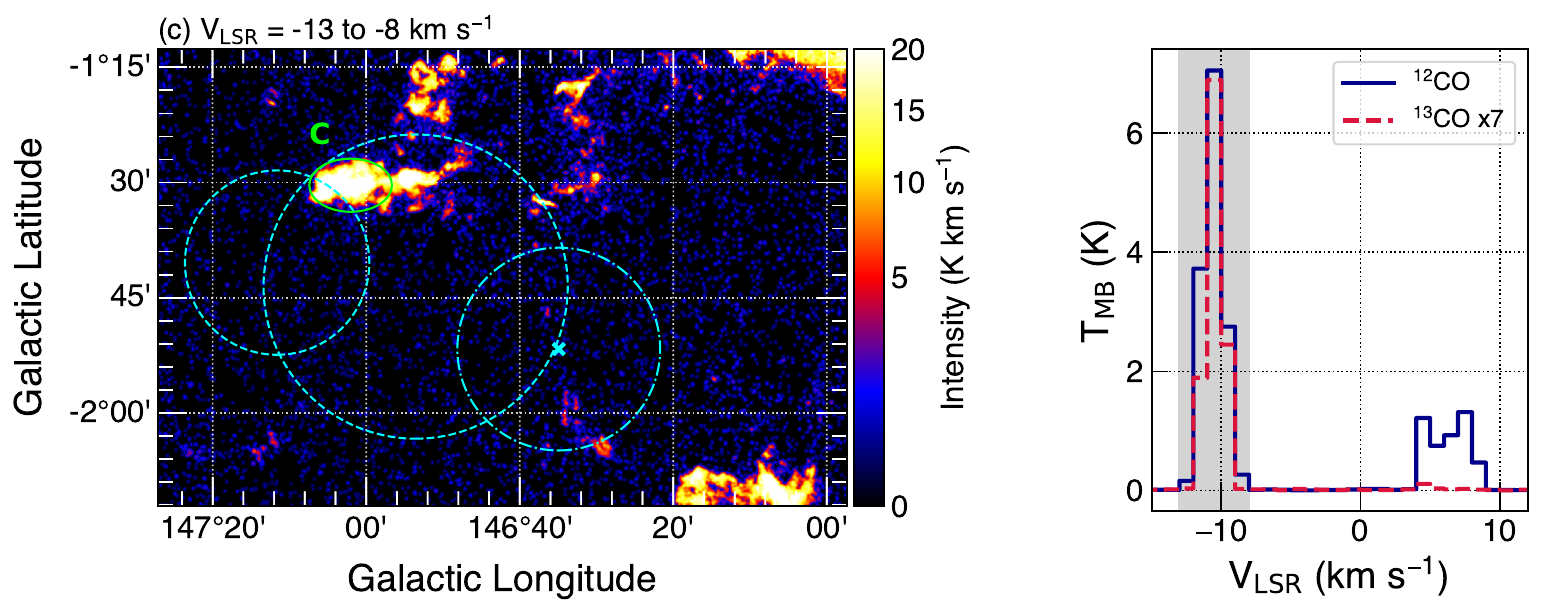}
  \includegraphics[width=0.6\linewidth]{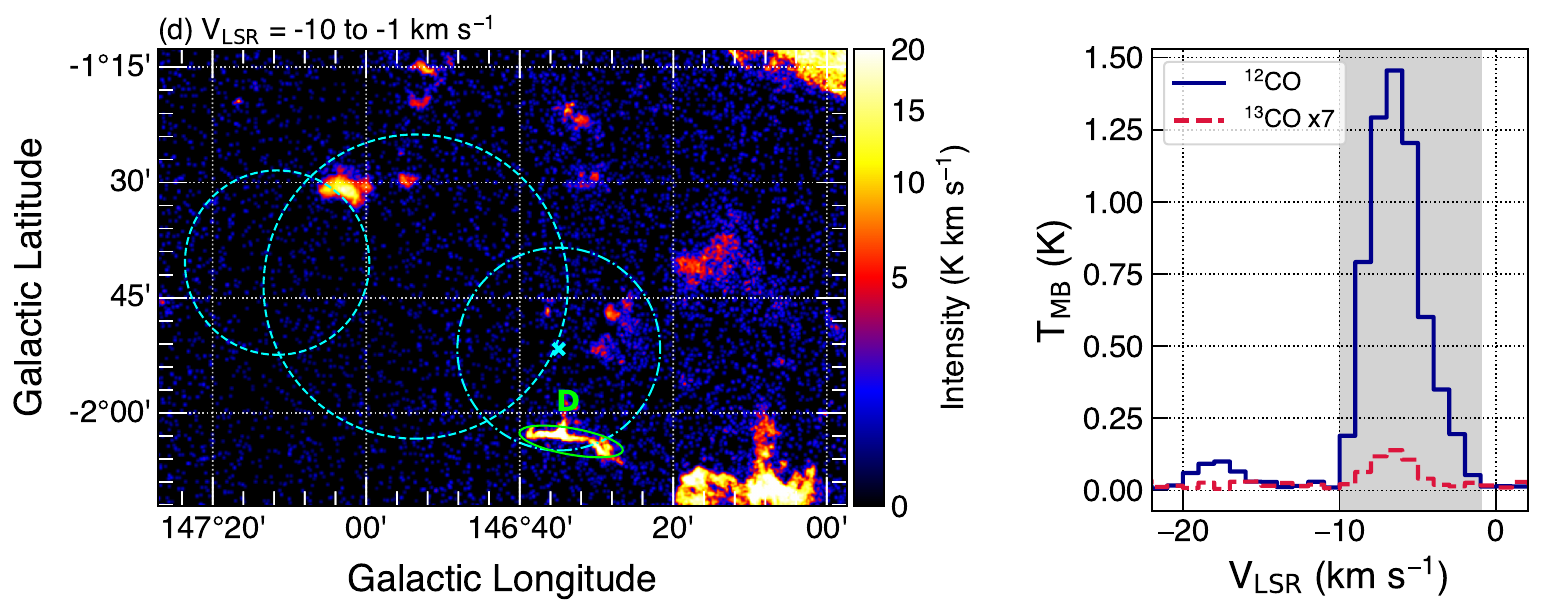}
  \includegraphics[width=0.6\linewidth]{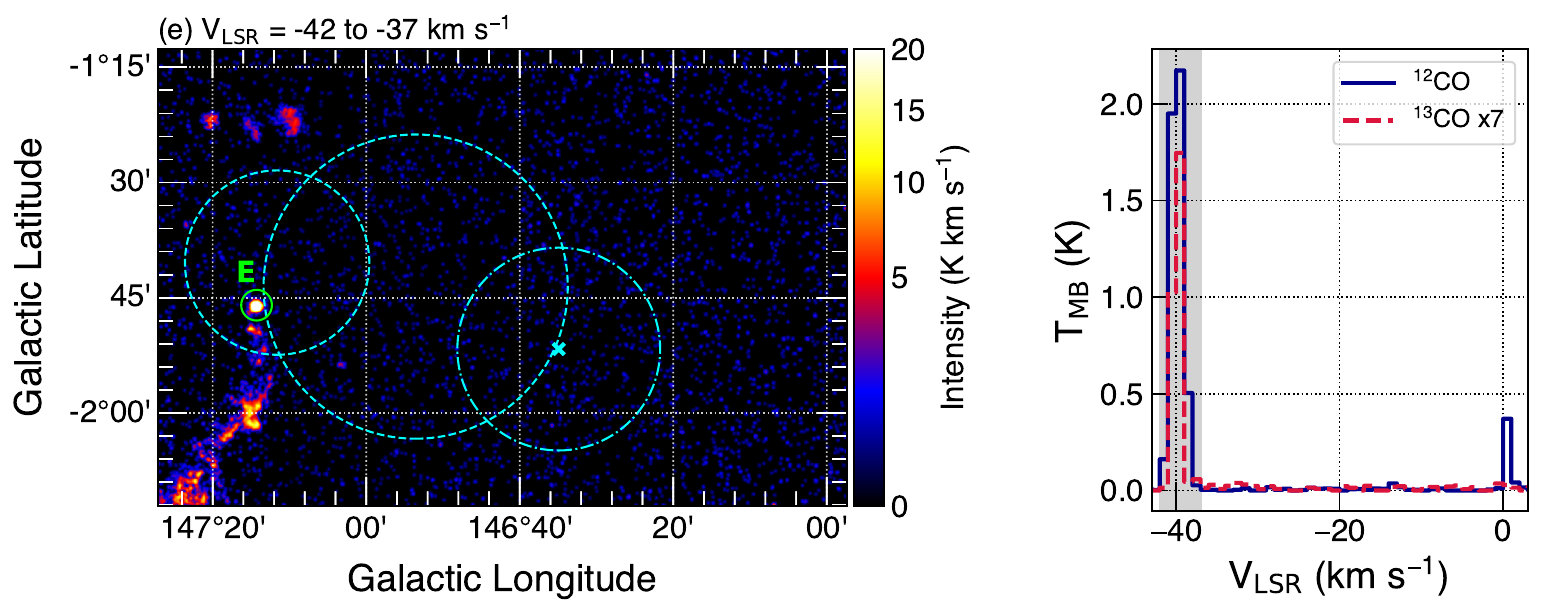}
 \end{center}
\caption{\COa\ map (left) and spectrum (right) of Clouds A to E (from top to bottom).
The grey shaded region in the spectrum indicates the velocity range that is used for the integration of the map. 
The gamma-ray emission of the first LHAASO catalogs is illustrated with cyan circles.
}
\label{fig:cloud}
\end{figure}

We found five molecular clouds within the gamma-ray emitting region, referred to as Clouds A--E.
\figref{fig:cloud} shows a velocity integrated map and spectrum extracted from each cloud.
The position and the size are summarized in \tabref{tab:cloud}.
We fit the velocity spectrum in \figref{fig:cloud} with a Gaussian model and show the best-fit values (i.e., the Gaussian center \Vlsr\ and full width at half maximum (FWHM) $\Delta$\Vlsr) in \tabref{tab:cloud}.

The \COc\ data were also analyzed in the same way. 
However, we could not find significant emission except for a tiny region, centered at ($\ell$, $b$) $=$ (147.0736\degr , $-$1.4945\degr) with a radius of 1.8 arcmin in the $-$13 to $-$8 \kms\ map.
The peak and averaged intensities within the circle are 3.1 K and 0.76 K, respectively, which are 11 and 2.8 times greater than the average RMS value of 0.27 K.
This \COc\ emission is located inside Cloud C and likely represents the ``core'' of that dense cloud.

\subsection{Physical parameters of molecular clouds}

In this subsection, we estimate physical parameters of the molecular clouds, including distance, column density, mass, and number density.

To estimate the distance to the clouds, we made use of a tool developed by \cite{wenger_kinematic_2018}\footnote{\url{https://www.treywenger.com/kd/index.php}}.
Given the obtained location and velocity of the molecular clouds, this tool calculates traditional kinematic distance by using the Galactic rotation curves \citep[e.g., ][]{brand_velocity_1993,reid_trigonometric_2014}. 
It also calculates Monte Carlo kinematic distance and its uncertainty from 10,000 Monte Carlo samples \citep{wenger_kinematic_2018}.
Since the direction to J0341 ($\ell=146.7$\degr) is near the Galactic anti center and oriented toward the outer Solar circle ($\approx$8 kpc from the Galactic center),
the kinematic distance can be poorly constrained, particularly for positive and small values of \Vlsr.
In \tabref{tab:cloud}, we show the distance derived by the Monte Carlo method \citep{wenger_kinematic_2018}, because the traditional kinematic distance cannot be derived in the J0341 region for Clouds A and B.

The column density, $N({{\rm H}_2})$, is estimated using two methods as follows.
The first method is to convert the \COa\ integrated intensity to $N({\mathrm{H}_2})$, assuming the X-factor of $X_{\rm CO} = 2 \times 10^{20} ~ {\rm cm}^{-2}~({\rm K ~ km ~s^{-1}})^{-1}$.
This factor has 10--40\% uncertainty at the Galactocentric radius of 8.5--12 kpc, which is derived for the position of the molecular clouds \citep{arimoto_co--h2_1996,bolatto_co--h2_2013,abe_prospects_2024}.
%
The second one is to calculate the column density assuming local thermodynamic equilibrium (LTE).
On the assumption that the \COa\ line is optically thick, the excitation temperature ($T_{\rm ex}$) is derived from the \COa\ data. 
With $T_{\rm ex}$ and the \COa\ intensity, the $^{13}$CO optical depth and column density, $N(^{13}{\rm CO})$, can be calculated. 
Then, $N(^{13}{\rm CO})$ is converted to $N({\rm H}_2)$ by multiplying the $^{13}$CO-to-H$_2$ factor of 7.7$\times 10^5$
\citep[see, e.g., ][for details]{wilson_tools_2009,de_la_fuente_detection_2023,sakemi_molecular_2023}.
It should be noted that the LTE method requires the use of the \COb\ data to derive $N(^{13}{\rm CO})$, therefore we applied it only to Clouds C and E, which the \COb\ emission is clearly detected (Figures~\ref{fig:map13} and \ref{fig:cloud}).
The difference between the two methods is a factor of 2--3 (\tabref{tab:cloud}).

The mass, $M$, of the cloud can be derived by
\begin{eqnarray}
    M = 2\mu m_{\rm H} d^2 \Omega  \sum_i \left[ N_i(\mathrm{H}_2)\right] ,    
\end{eqnarray}
where 
$\mu$ ($\approx$1.4) is mean molecular weight,
$m_{\rm H}$ is mass of hydrogen, 
$d$ is the distance, 
$\Omega$ is an angular area of a grid space (i.e., 7.5$\times$7.5 arcsec$^2$),
and $i$ indicates each pixel.
\tabref{tab:cloud} summarizes the estimated mass, as well as a number density, $n$, derived by assuming a sphere.
Note that Virial mass \citep{garay_massive_1999} is much larger than the estimated mass for all of the detected molecular clouds. 
In \tabref{tab:cloud}, the uncertainties of \Vlsr\ and $\Delta$\Vlsr\ are statistical ones when fitting the velocity spectra,
that of the distance is taken from the estimation method of the Monte Carlo kinematic distance \citep{wenger_kinematic_2018},
ant those of $N$, $M$, and $n$ arise from the RMS.
It should be noted that the systematic uncertainty on the velocity arises from its resolution, that is 1~\kms\ in our observations.

\begin{table*}[ht!]
\begin{center}
  \caption{Physical parameters of molecular clouds in the J0341 region}
  \begin{tabular}{ccccccccccccccc}
      \hline
 Cloud &     $\ell$ &    $b$ &  $V_{\rm LSR}$ &  $\Delta V_{\rm LSR} ^*$ &  $d$ &  Radius &  $T_{\rm ex}$ &  $N(\mathrm{H}_2) ^\dagger$ &     $M(\mathrm{H}_2)$ &      $M_2 (\mathrm{H}_2) ^\S$ & $n(\mathrm{H}_2) ^\ddagger$  \\ 
  &     (deg) &    (deg) &  (\kms) &  (\kms) &  (kpc) &  (pc) & (K) & (10$^{21}$ \columnd) &     (\Msun) &   (\Msun) &  (\cc)  \\ 
\hline
A & 147.0 & -1.74 & 1.33 $\pm$ 0.01 & 1.21 $\pm$ 0.02 & $<$0.4 & 0.52 & 4.1 $\pm$ 3.9 & 0.28 $\pm$ 0.2 & 5.4 $\pm$ 3.6 & --- & 370 $\pm$ 240  \\ 
B & 146.9 & -1.46 & 4.94 $\pm$ 0.17 & 3.46 $\pm$ 0.41 & $<$0.3 & 0.92 & 7.3 $\pm$ 4.0 & 1.2 $\pm$ 0.2 & 25 $\pm$ 4 & --- & 310 $\pm$ 50  \\ 
C & 147.0 & -1.51 & -10.59 $\pm$ 0.01 & 1.85 $\pm$ 0.02 & 0.9 $\pm$ 0.4 & 1.6 & 10 $\pm$ 4 & 2.8 $\pm$ 0.2 & 310 $\pm$ 130 & 150 $\pm$ 70 & 780 $\pm$ 320  \\ 
D & 146.6 & -2.06 & -6.56 $\pm$ 0.06 & 3.87 $\pm$ 0.13 & 0.5 $\pm$ 0.4 & 0.99 & 4.5 $\pm$ 4.1 & 1.2 $\pm$ 0.2 & 21 $\pm$ 18 & --- & 210 $\pm$ 170  \\ 
E & 147.2 & -1.77 & -39.88 $\pm$ 0.01 & 1.77 $\pm$ 0.03 & 4 $\pm$ 1 & 1.2 & 12 $\pm$ 4 & 3.6 $\pm$ 0.2 & 340 $\pm$ 90 & 130 $\pm$ 40 & 2100 $\pm$ 530  \\ 
\hline
    \end{tabular}
    \label{tab:cloud}
\end{center}
\tablecomments{\\
$*$ $\Delta V_{\rm LSR}$ indicates FWHM.  \\ 
$\dagger$ $N(\mathrm{H}_2)$ is averaged within the region.  \\ 
$\S$ $M_2$ is a mass derived assuming LTE.  \\ 
$\ddagger$ $n$ is derived assuming a sphere. \\
}
\end{table*}

\subsection{Summary} \label{sec:cloud}
As summarized in \tabref{tab:cloud}, we found that most of the molecular clouds within the gamma-ray emitting region are nearby ($d \lesssim$ 1 kpc), relatively small ($\lesssim$ a few pc), and low in mass ($M \lesssim $ a few hundreds of solar masses).
Descriptions of each cloud are provided below (see Appendix~\ref{sec:outside} for clouds outside the gamma-ray region\footnote{The gamma-ray radius presented here indicates the 39\% containment radius in the 2D Gaussian model. Thus, the molecular clouds outside $r_{39}$ can potentially act as targets of proton-proton interactions. See also \figref{fig:grid} for the 68\% containment radius overlaid the molecular cloud map.}): 
\begin{itemize}
    \item Cloud A (half-shell): as mentioned in \cite{cao_discovery_2021}, ``half-shell'' structure is located almost at the center of 1LHAASO J0343$+$5254u with WCDA (previous LHAASO J0341$+$5258 with KM2A). It is nearby ($d < 0.4$ kpc), and its mass is 5.4 \Msun. Although the projected image looks like a half shell, the velocity-position map does not show any hint of an expanding feature. Observations with the better velocity resolution might be needed to confirm its property as an expanding shell. 
    \item Clouds B and C (dense core): Cloud C is detected in all \COa, \COb, and \COc\ lines, indicating it is a dense core of the cloud. Both Clouds B and C are distributed in two velocities of $-$10 \kms\ and 5 \kms. The association of these two clouds is not clear. 
    \item Cloud D (filament-like): This cloud is located near the outer edge of the gamma-ray source LHAASO J0339$+$5307 with KM2A. 
    The velocity width is relatively broad, $\Delta V \sim$4 \kms, among the clouds in the J0341 region. 
    \item Cloud E (point-like): Cloud E has a compact, point-like structure, although it is extended compared to the angular resolution of 15\arcsec\ of \nobeyama. This source has optical and infrared counterparts, referred to as IRAS~03392$+$5239 \citep{wouterloot_iras_1989}. It is likely an evolved star, such as an asymptotic giant branch (AGB) star, surrounded by a CO envelope. However, the progenitor star of an AGB star has typically a mass of $\lesssim$10 \Msun, which contradicts the obtained mass of 340~\Msun\ in Cloud E. The mass estimated by our CO data could be contaminated by CO gas that is not directly associated with the AGB star. High-resolution observations (e.g., by ALMA) may help clarify the association. Alternatively, Cloud E may represent a compact cloud clump or a core of molecular cloud \citep[e.g., ][]{mckee_theory_2007}. A possibility of particle acceleration at Cloud E is discussed in \secref{sec:discussion}.
\end{itemize}

\section{\HI\ and continuum observation} \label{sec:CGPS}

In addition to the CO data with \nobeyama\, 
we analyzed the archival data of atomic hydrogen (\HI) 21-cm line and continuum emission.
These data were retrieved from the Canadian Galactic Plane Survey (CGPS)\footnote{\url{https://www.cadc-ccda.hia-iha.nrc-cnrc.gc.ca/en/search/}}.
We made use of the \HI\ observations and continuum observations at 408 and 1420 MHz, covering the J0341 region, from the Dominion Radio Astrophysical Observatory (DRAO) archives \citep{taylor_canadian_2003}.
These observations were performed in 2003--2004.
The \HI\ data have an angular resolution of 18\arcsec\ and a velocity range from $-$165 \kms\ to 59 \kms\ with a grid of 0.82 \kms. 
For the continuum radio data, the spatial resolution is 18\arcsec\ at the 1420 MHz (Stokes I) band and 54\arcsec at the 408 MHz band.

\subsection{\HI\ cloud}

\begin{figure*}[ht!]
 \begin{center}
  \includegraphics[width=0.9\linewidth]{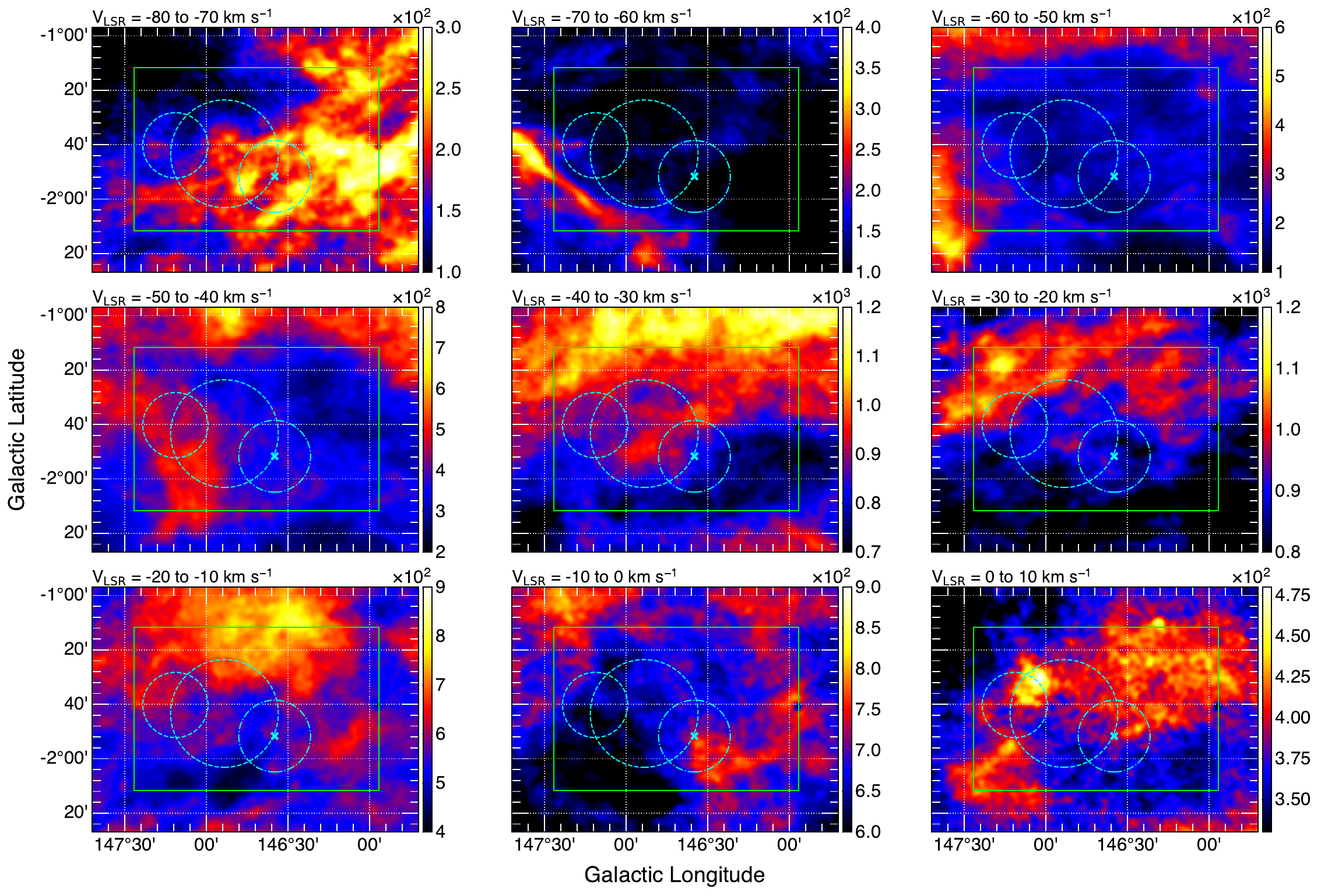}
 \end{center}
\caption{
\HI\ intensity maps at \Vlsr\ of $-$80 to 10 \kms\ with a grid of 10~\kms,
overlaid with the gamma-ray emission of the first LHAASO catalogs in cyan circles and RoI in green box.
The intensity is shown in the units of K~km~s$^{-1}$.
}
\label{fig:HI_grid}
\end{figure*}

\begin{figure*}[ht!]
 \begin{center}
  \includegraphics[width=0.6\linewidth]{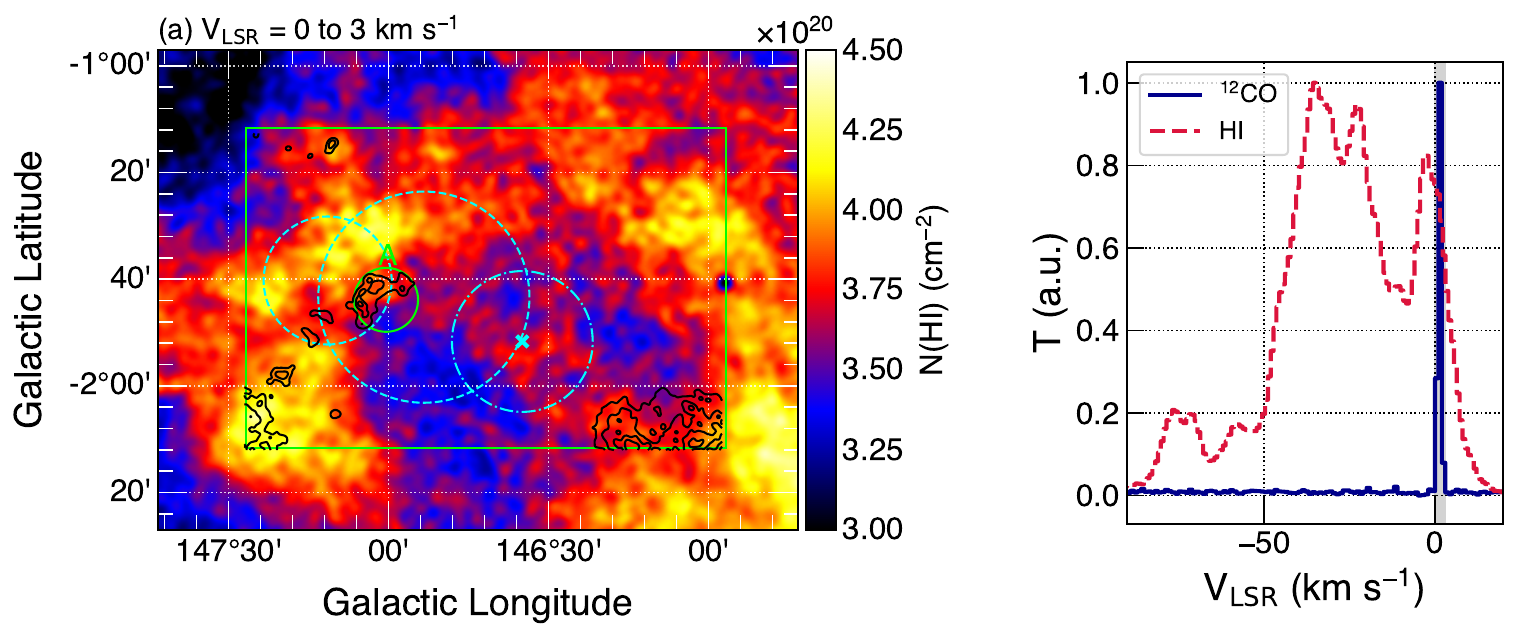}
  \includegraphics[width=0.6\linewidth]{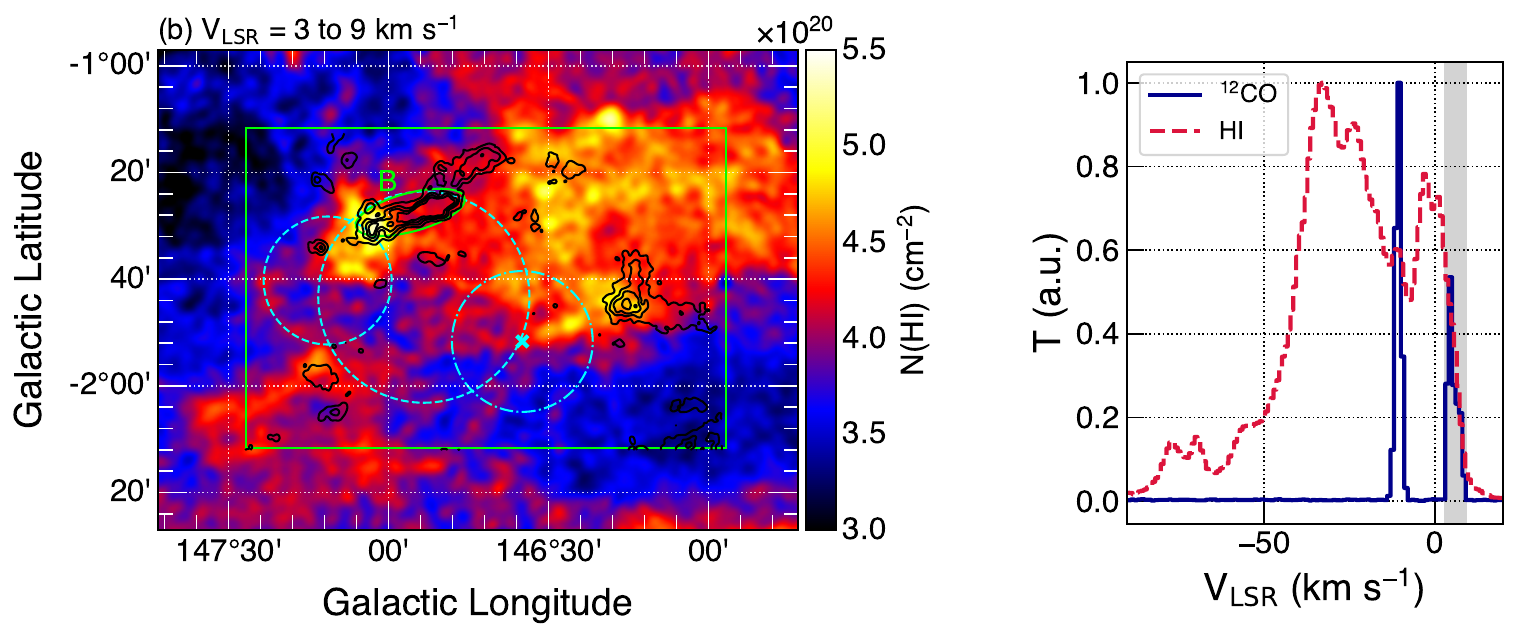}
  \includegraphics[width=0.6\linewidth]{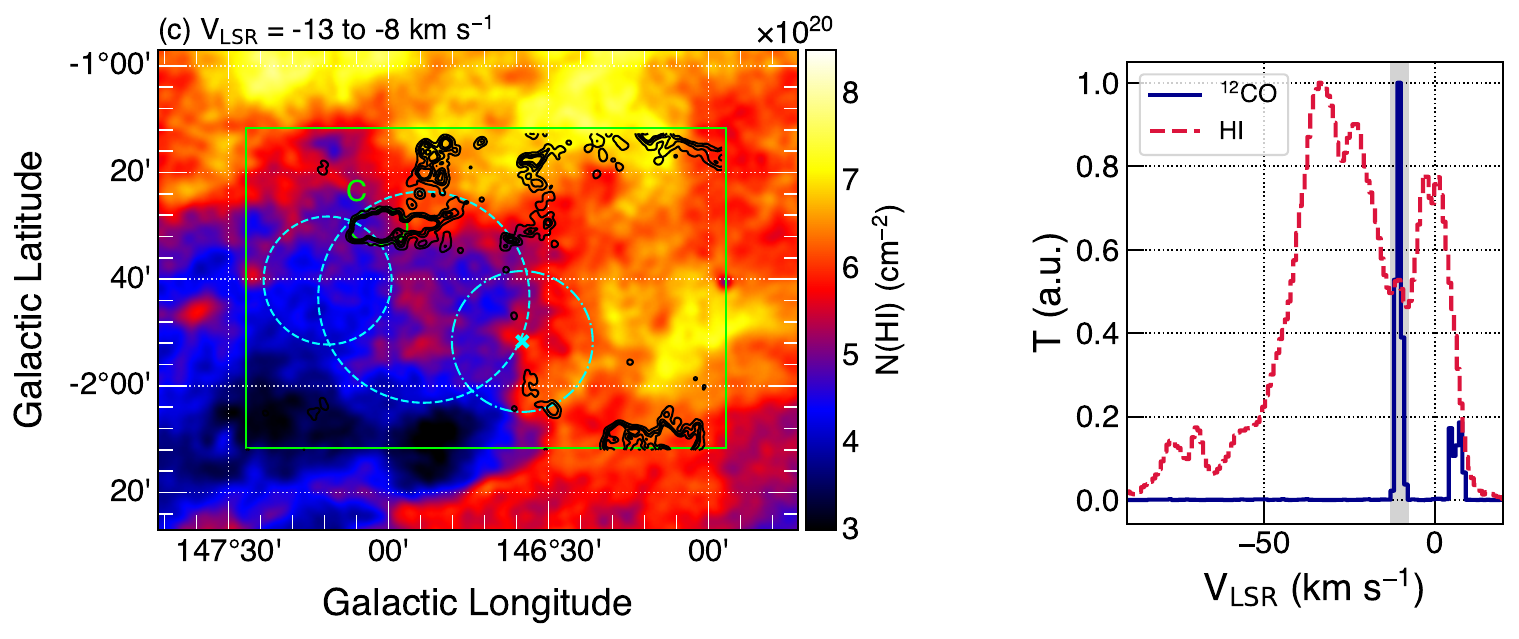}
  \includegraphics[width=0.6\linewidth]{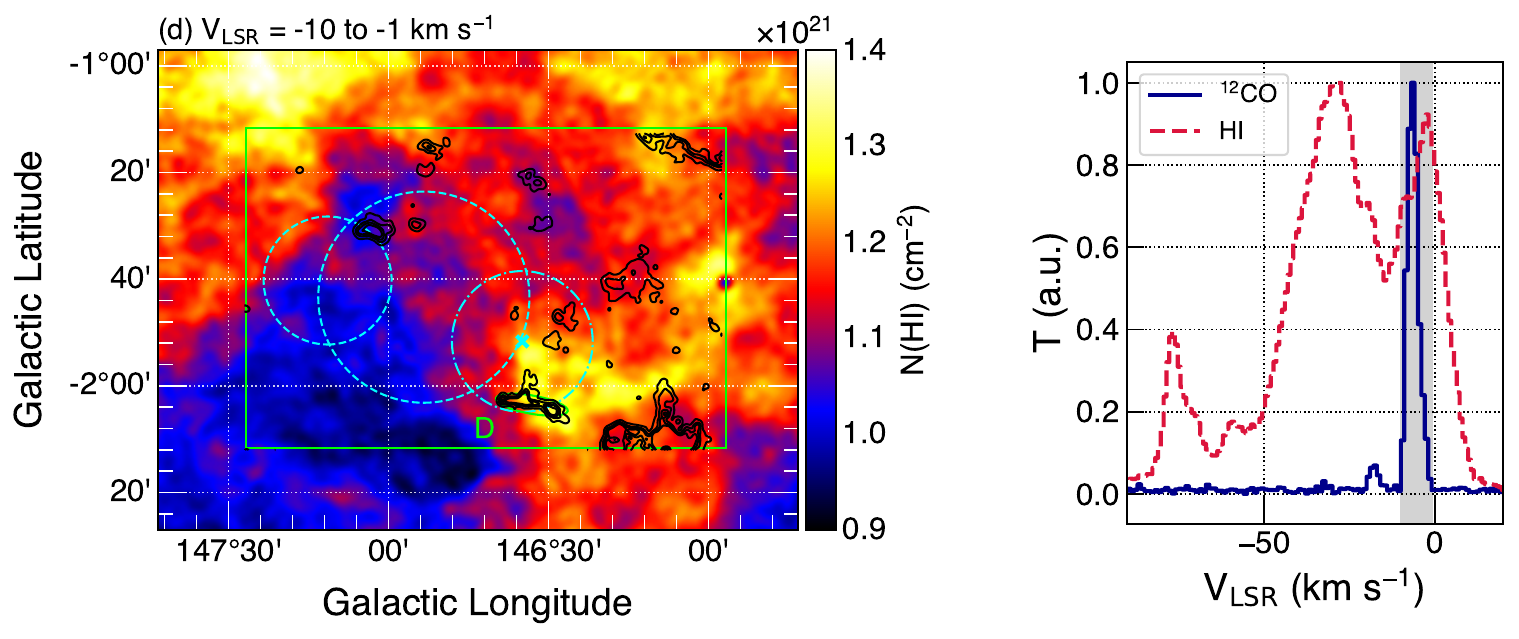}
  \includegraphics[width=0.6\linewidth]{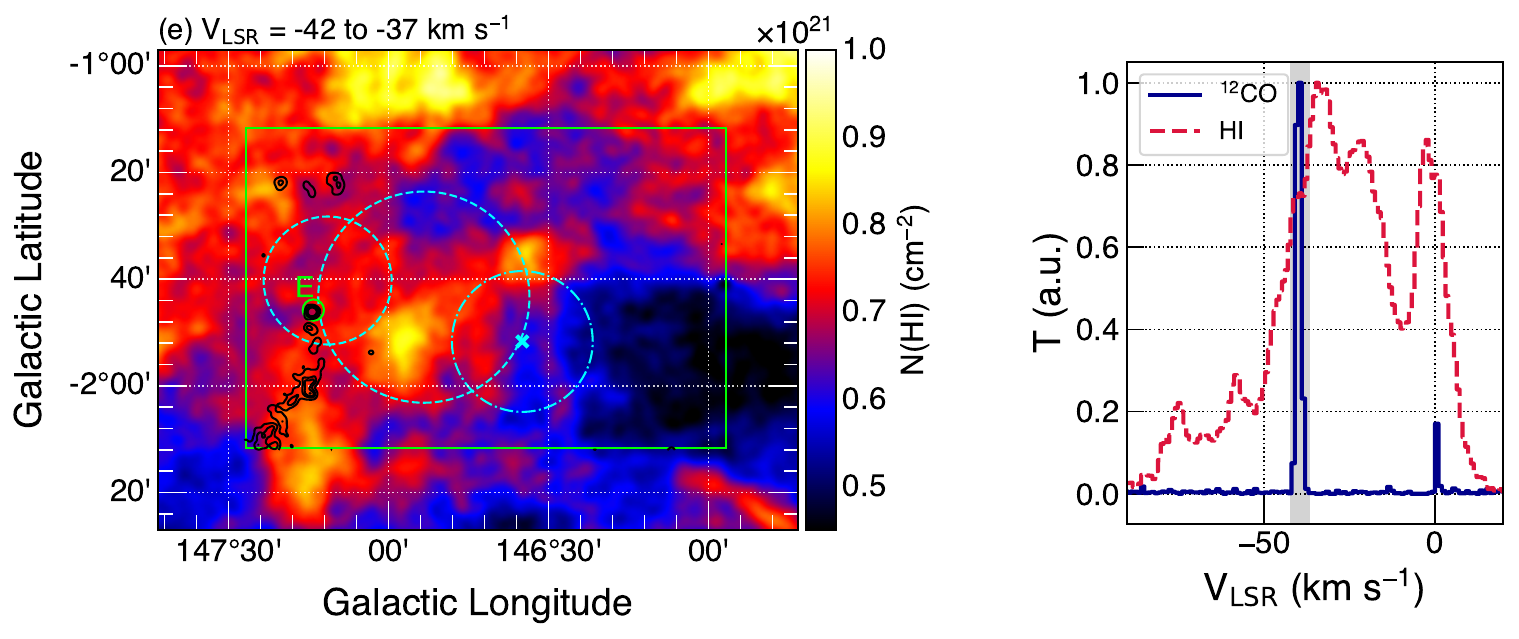}
 \end{center}
\caption{
Left: \HI\ map in the same velocity ranges of Clouds A--E, overlaid with the gamma-ray emission of the first LHAASO catalogs in cyan circles, RoI in green box, and the \COa\ contour in black lines.
Right:  \HI\ and \COa\ spectra.
}
\label{fig:HIcloud}
\end{figure*}

In \figref{fig:HI_grid}, we show the \HI\ map integrated in the velocity range from $-$80 \kms\ to 10 \kms\ with a grid of 10 \kms.
The \HI\ emission is diffusively extended over our RoI, and there is no apparent counterpart of the gamma-ray emission observed by LHAASO and the molecular clouds by \nobeyama.
Assuming the \HI-to-H conversion factor of 1.823$\times 10^{18} ~{\rm cm}^{-2}~({\rm K ~ km ~ s^{-1}})^{-1} $ \citep{dickey_h_1990}, we calculated the column density $N$(\HI).
\figref{fig:HIcloud} compares the map and spectrum of \HI\ with those of \COa.
Although the spectral shape is different between the \HI\ and CO emission,
we calculated $N$(\HI) (and mass in turn) extracted from the same velocity range with that of CO.
%
The proton column density, $N({\rm H})$ (and also the number density $n({\rm H})$), is derived by
\begin{eqnarray}
    N({\rm H}) = 2 \times N({\rm H}_2) + N({\rm HI}) .
\end{eqnarray}
The total mass and number density of protons are summarized in \tabref{tab:HI}.

If we derive the \HI\ amount from the same region and velocity range of the CO data,
the obtained mass and density are comparable to those of the molecular clouds in Clouds A and D and smaller in the rest (\tabref{tab:HI}).
The total proton number density, $n(\mathrm{H}_2 + {\rm HI})$, is estimated to be $\sim 10^3$~\cc.

\begin{table*}[ht!]
\begin{center}
    \caption{Comparison of the CO and \HI\ clouds in J0341}
    \begin{tabular}{ccccccccc}
      \hline
Cloud & $N(\mathrm{H}_2) ^*$  &  $N({\rm HI}) ^*$  & $M({\mathrm{H}_2})$ &  $M({\rm HI})$ & $M(\mathrm{H}_2 + {\rm HI})$ & $n(\mathrm{H}_2)$ & $n({\rm HI})$ & $n(\mathrm{H}_2 + {\rm HI})$ \\
 & (10$^{21}$ \columnd) &  (10$^{21}$ \columnd) & (\Msun) &  (\Msun) & (\Msun) & (\cc) & (\cc)  & (\cc) \\
\hline
A & 0.28 $\pm$ 0.20 & 0.37 $\pm$ 0.01 & 5.4 $\pm$ 3.6 & 7.2 $\pm$ 0.1 & 13 $\pm$ 4    & 370 $\pm$ 240  & 480 $\pm$ 8   & 1200 $\pm$ 480  \\ 
B & 1.2 $\pm$ 0.2 & 0.44 $\pm$ 0.01  & 25 $\pm$  4   & 9.0 $\pm$ 0.1 & 34 $\pm$ 4    & 310 $\pm$ 50   & 110 $\pm$ 2   & 740  $\pm$ 100  \\ 
C & 2.8 $\pm$ 0.2 & 0.50 $\pm$ 0.01  & 310 $\pm$ 130 & 55 $\pm$ 22   & 360 $\pm$ 130 & 780 $\pm$ 320  & 140 $\pm$ 60  & 1700 $\pm$ 640  \\ 
D & 1.2 $\pm$ 0.2 & 1.40 $\pm$ 0.01  & 21 $\pm$ 18   & 23 $\pm$ 19   & 45 $\pm$ 26   & 210 $\pm$ 170  & 230 $\pm$ 180 & 650  $\pm$ 390  \\ 
E & 3.6 $\pm$ 0.2 & 0.59 $\pm$ 0.01  & 340 $\pm$ 90  & 56 $\pm$ 14   & 400 $\pm$ 88  & 2100 $\pm$ 530 & 350 $\pm$ 90  & 4500 $\pm$ 1100  \\ 
\hline
    \end{tabular}
    \label{tab:HI}
\end{center}
\tablecomments{ \\
$*$ The column density, $N$, is the averaged value within the region, and the spatial grid size is 7.5\arcsec\ and 18\arcsec\ for the CO and \HI\ data, respectively. \\
}
\end{table*}

\subsection{Continuum}

\begin{figure*}[ht!]
 \begin{center}
  \includegraphics[width=0.45\linewidth]{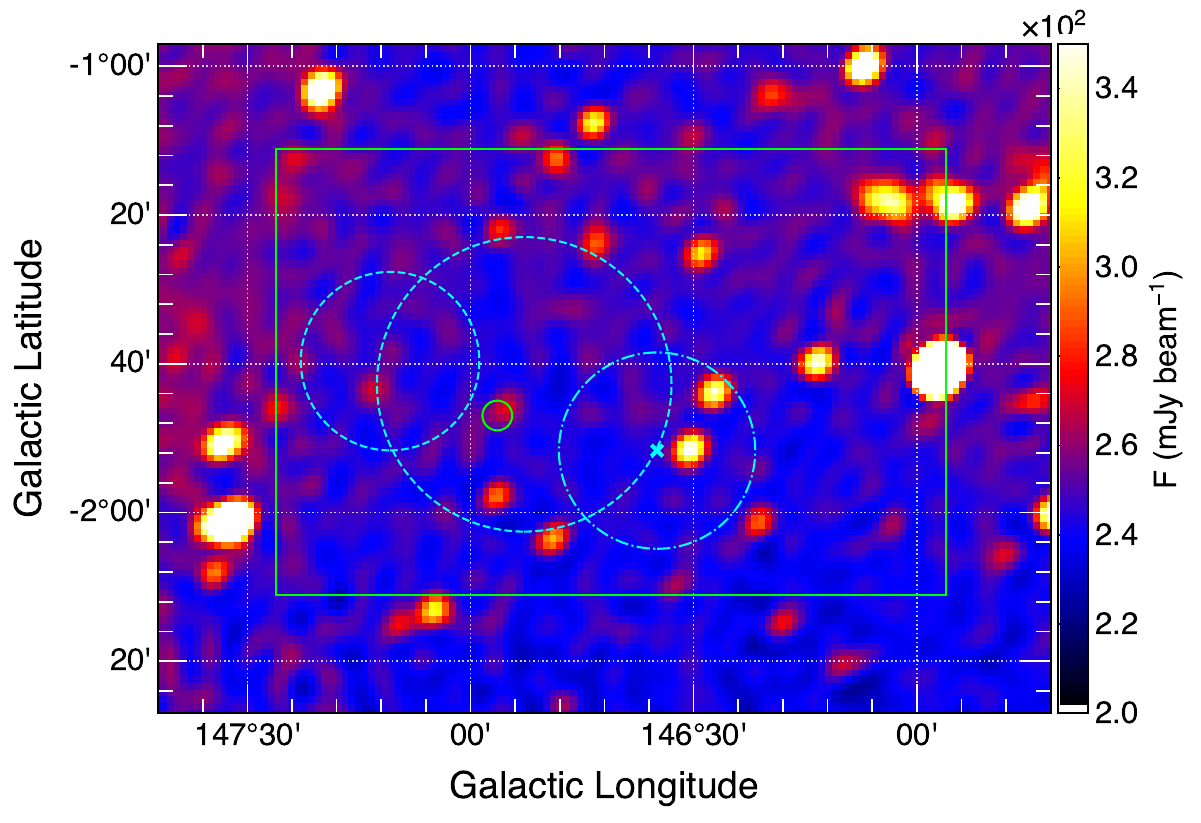}
  \includegraphics[width=0.45\linewidth]{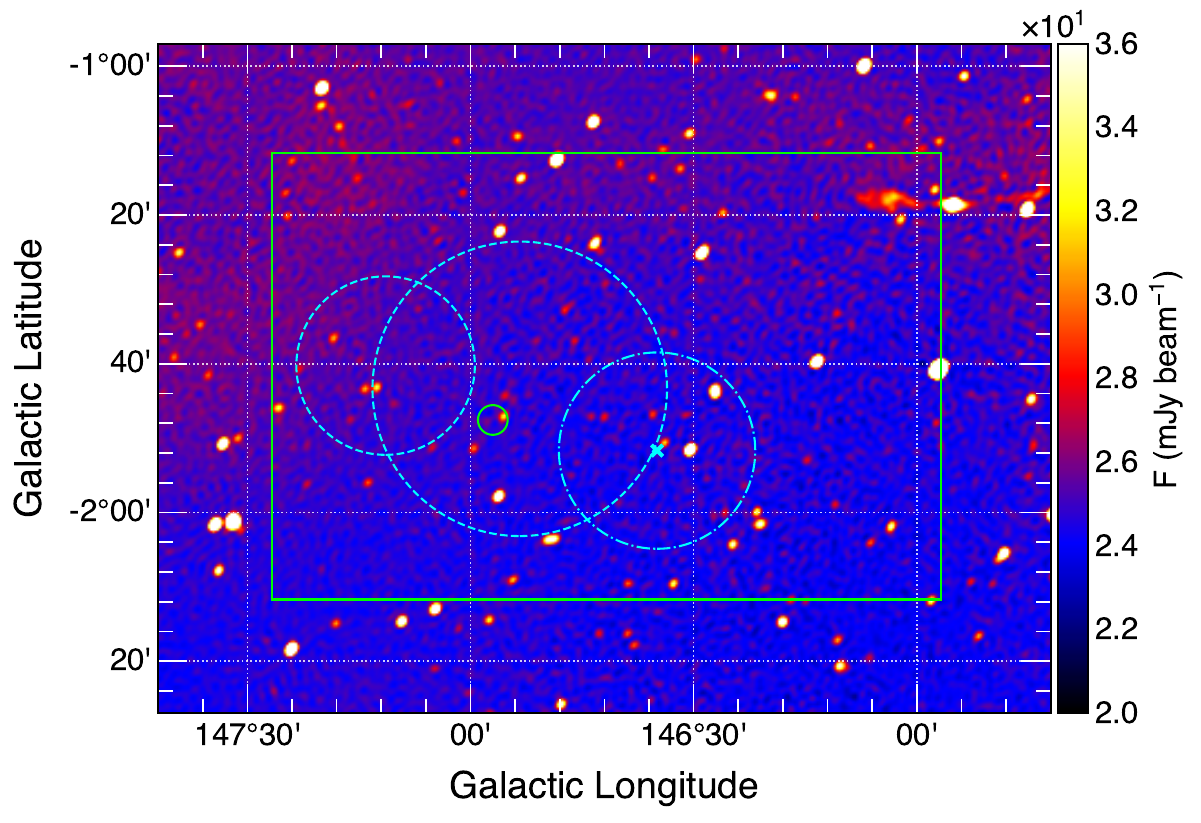}
 \end{center}
\caption{
Radio continuum map at 408 MHz (left) and 1420 MHz (right),
overlaid with the gamma-ray emission of the first LHAASO catalogs in cyan lines and RoI in green box.
The green circle indicates the position of the X-ray extended source detected by \xmm\ \citep{dikerby2024}.
}
\label{fig:continuum}
\end{figure*}

\figref{fig:continuum} shows the continuum maps at 408 MHz and 1420 MHz.
The flux density, $F$, can be derived as
\begin{eqnarray}
    \left( \frac{F}{{\rm mJy~beam}^{-1}} \right) = 8.2\times 10^{-4} \left(\frac{T}{\rm K}\right)  \left(\frac{\nu}{\rm GHz}\right)^2  \left(\frac{\theta_{\rm major} \times \theta_{\rm minor}}{{\rm arcsec}^2}\right) ,
\end{eqnarray}
where $T$ and $\nu$ are a brightness temperature and frequency, respectively.
A synthesized beam size ($\theta_{\rm major} \times \theta_{\rm minor}$) can be calculated as
2.8\arcmin\ $\times$ 2.8\arcmin\ $\times {\rm cosec}\delta$  at 408 MHz and
49\arcsec\ $\times$ 49\arcsec\ $\times {\rm cosec}\delta$ at 1420 MHz, where $\delta$ is the declination \citep{taylor_canadian_2003}.
With $\delta$ of 52.97\degr\ in the J0341 region, the beam size is 2.8\arcmin\ $\times$ 3.5\arcmin\ and 49\arcsec\ $\times$ 61\arcsec\ at 408 MHz and 1420 MHz, respectively.
The RMS noise level is 3.8 (408 MHz) and 0.27 (1420 MHz) mJy~beam$^{-1}$ \citep{taylor_canadian_2003}.
There is no diffuse emission corresponding to the gamma rays but some point-like sources.
\tabref{tab:continuum} shows the radio continuum flux, extracted from the gamma-ray emitting regions detected by LHAASO.
Since the background is not taken into account, the flux should be considered an upper limit.
Note that excluding the point-like sources does not largely affect the obtained flux. 

\begin{table}[ht!]
\begin{center}
  \caption{Radio continuum flux in J0341}
  \begin{tabular}{ccccc}
      \hline
 Name & Instrument  &   $F_{\rm 408~ MHz}$ &   $F_{\rm 1420~ MHz}$  \\ 
      &   & (erg cm$^{-2}$ s$^{-1}$) & (erg cm$^{-2}$ s$^{-1}$) \\
\hline
1LHAASO J0339+5307  & LHAASO KM2A & 5.6$\times 10^{-14}$ & 2.3$\times 10^{-13}$  \\
1LHAASO J0343+5254u & LHAASO KM2A & 4.6$\times 10^{-14}$ & 2.0$\times 10^{-13}$  \\
1LHAASO J0343+5254u & LHAASO WCDA & 1.3$\times 10^{-13}$ & 5.2$\times 10^{-13}$  \\
1LHAASO total$^*$ &  & 2.0$\times 10^{-13}$ & 8.2$\times 10^{-13}$  \\
\hline 
New extended X-ray  & \xmm & 1.2$\times 10^{-15}$ & 4.3$\times 10^{-15}$ \\
\hline
    \end{tabular}
    \label{tab:continuum}
    \end{center}
\tablecomments{ 
$^*$ 1LHAASO J0339$+$5307 and 1LHAASO J0343$+$5254u are merged. \\ 
}
\end{table}

\section{Discussion}  \label{sec:discussion}

\begin{figure}[ht!]
    \centering
    \includegraphics[width=0.8\linewidth]{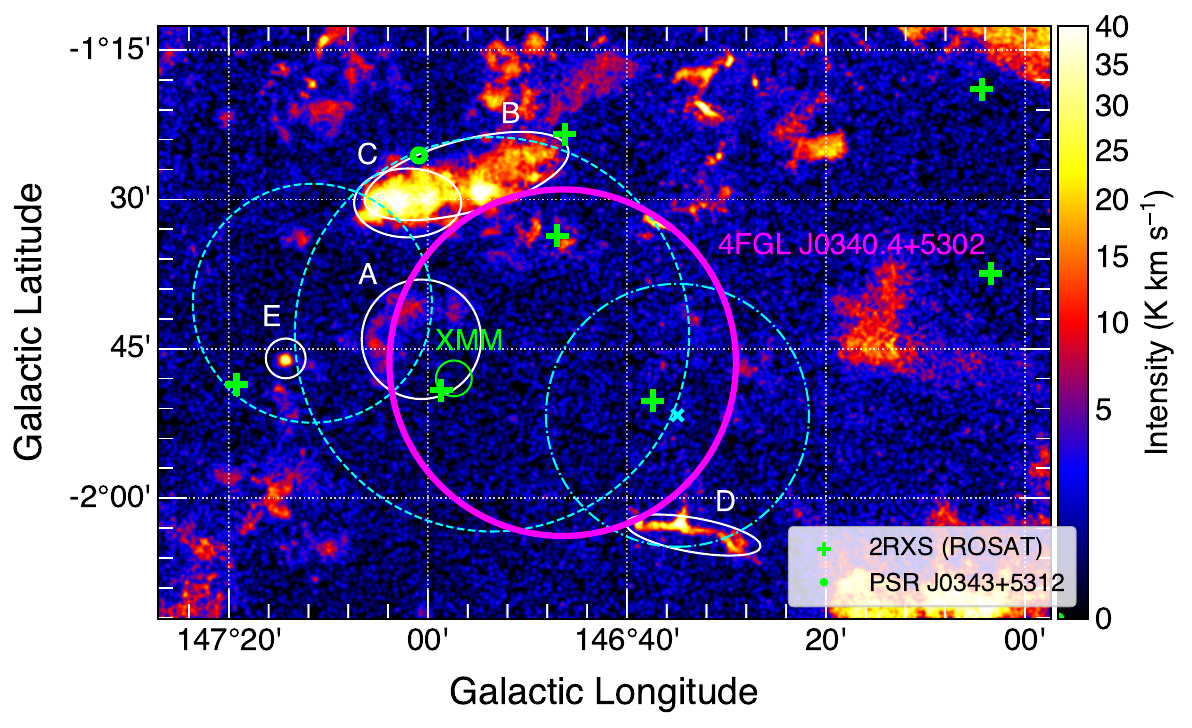}
    \caption{
    The \COa\ map with \Vlsr\ = $-$40 to 10 \kms, overlaid with the gamma-ray emission of the first LHAASO catalogs (cyan circles), known counterparts of PSR J0343$+$5312 (a green small circle), 4FGL J0340.4$+$5302 (a magenta circle), \rosat\ sources from the 2RXS catalog (green crosses), and newly detected \xmm\ source (the green circle; \cite{dikerby2024}). 
    Clouds A--E are marked in white.
    }
    \label{fig:counterpart}
\end{figure}

\figref{fig:counterpart} shows the known counterparts within the RoI, including a pulsar (PSR J0343$+$5312), an unidentified \lat\ GeV gamma-ray source (4FGL J0340.4$+$5302), \rosat\ X-ray sources from the 2RXS catalog. 
PSR J0343$+$5312 is unlikely relevant to the UHE gamma-ray emission, given its low power of $\sim 10^{31}$ erg~s$^{-1}$.
We also show the position of a newly detected \xmm\ source, which is likely a pulsar wind nebula (PWN) \citep{dikerby2024}.
The new \xmm\ source is located near Cloud A, thus if they are associated with each other, the X-ray source could be at less than 0.4~kpc.
The flux of the radio continuum emission from this X-ray source is $\lesssim  10^{-15}$~\eflux\ (\tabref{tab:continuum}).
It should be noted that the ROSAT point source (2RXS~J034125.8$+$525530), located near the edge of Cloud A and the \xmm\ extended source, did not appear in the new \xmm\ data \citep{dikerby2024}, suggesting that it is a transient source and not related to the extended emission.
The PWN scenario, including a detailed leptonic spectral energy distribution modeling, is presented in a separate publication \citep{dikerby2024}.

If the gamma-ray emission is hadronic, a total proton energy ($W_p$) can be calculated as
$W_p = 4\pi d^2 F_\gamma t_{\pi^0} \eta^{-1}$,
where $F_\gamma$, $t_{\pi^0} \ (\approx 1.5\times10^{15} n^{-1}~ {\rm s})$, and $\eta \ (\approx 1.5$--2) denote the gamma-ray flux, a time scale of proton-proton interactions, and a parameter to take into account the gamma-ray production in interactions involving nuclei of both CRs and interstellar medium \citep{Kafexhiu2014}, respectively.
Adopting the gamma-ray flux of $2.4 \times 10^{-12}$~\eflux\  \citep{cao_first_2024},
\begin{equation}
    W_p (E>1~{\rm TeV}) \sim 3 \times 10^{45} \left( \frac{d}{1~{\rm kpc}}\right)^2 \left( \frac{n}{100~{\rm cm}^{-3}}\right)^{-1} ~{\rm erg} .
    \label{eq:energetics}
\end{equation}
Note that the proton spectrum was obtained to be a cutoff power-law model with an index of $\leq2$ and a cutoff energy of 200--300 TeV in the hadronic case \citep{cao_discovery_2021,kar_ultrahigh-energy_2022,de_sarkar_dissecting_2023}.
Although the molecular clouds detected by \nobeyama\ (Clouds A--E) do not have exact counterparts of the known sources,
the observations provided important parameters such as the distance and number density.
Some of the known sources and the gamma-ray emission may be located at the same distance as Clouds A--E.
If so, the distance would be $\lesssim$1 kpc or $\sim$4 kpc.
The total number density of protons in Clouds A--E is $n \sim$ 700--5000~\cc\ (\tabref{tab:HI}).
Thus, the proton energy could be $\lesssim10^{45}$~erg, inferred from Equation~\ref{eq:energetics}.
This is much smaller than the typical SNR value, where the explosion energy is $\sim 10^{51}$~erg, and the conversion efficiency from the explosion energy to CR acceleration is $\sim$1--10\%.
If the gamma-ray radiation is of hadronic origin, it might originate from molecular clouds illuminated by escaping CRs, rather than any hadronic acceleration site.
The narrow velocity width of 1--4~\kms\ of the detected clouds may also support the idea that dynamical power, which could be expected for CR accelerators such SNRs, is not injected.
Further observations with the better velocity resolution and/or other transition line emissions would be helpful.
As inferred from our analysis, the distance to most of the clouds could be smaller than 1 kpc, and thus there may exist hadronic accelerators outside our RoI in this paper.
%
In the vicinity of J0341, there are roughly 10 SNRs within 20\degr\ \citep{SNRcat}. This corresponds to $<$ 360 pc for $d < 1$~kpc and $<$ 1.2 kyr for protons to travel, which is consistent with the assumptions in \cite{mitchell_using_2021}. 
We found three of them (G156.2$+$05.7, G160.9$+$02.6, and G166.0$+$04.3) might have potential contribution to the J0341 region because they are relatively young, shell-like, and likely detected in the GeV gamma-ray band.

In this paper, we studied the distribution of CO and \HI\ to investigate the gas in the gamma-ray emitting region.
Additionally, there exists the so-called ``dark gas'', which cannot be traced by the CO and \HI\ line emissions.
It has been shown that the amount of the dark gas would be likely $\sim$0.1--5 times that of H$_2$ and \HI, depending on the total mass \citep[e.g., ][]{grenier_unveiling_2005,grenier_nine_2015,mizuno_gas_2022}. 
The detection of several molecular clouds by the \nobeyama\ suggests that these clouds, which may be surrounded by the dark gas, could be potential target gas to produce hadronic gamma rays,
even though their angular sizes are smaller than those of the gamma-ray emission.
Taking into account the dark cloud, the total mass and number density might be several times larger, and thus $W_p$ may become smaller than 10$^{45}$ erg in the hadronic scenario.

Cloud E, which is likely a CO envelope of an AGB star (IRAS 03392$+$5239 \citep{wouterloot_iras_1989}), is located at nearly the center of 1LHAASO J0343$+$5254 (KM2A), as shown in the bottom panel of \figref{fig:cloud}.
Remarkably, there is a tail-like structure that is likely launched from the compact CO object and that might be a bow shock \cite[e.g., ][]{cox_far-infrared_2012}.
Although the association of Cloud E with the AGB star is unclear as mentioned in \secref{sec:cloud},
we now explore the possibility of particle acceleration at AGB stars to test whether \revise{Cloud E} itself can act as a CR accelerator.
Since the size ($D$) of Cloud E is 2 arcmin in diameter,
corresponding to 2.3 pc at $d=4$~kpc, and the velocity width ($\Delta V$) is 1.8~\kms\ (\tabref{tab:cloud}), a dynamical time can be estimated as $t_{\rm dyn} = D/\Delta V = $1.3~Myr.
Given its mass of 340~\Msun\ (\tabref{tab:cloud}), a mass loss rate is calculated as $\dot{M} = M/t_{\rm dyn} = 2.6 \times 10^{-4}$~\Msun~yr$^{-1}$.
Applying the same calculation to the tail-like structure, we obtained a size of 20 pc, a mass of 950 \Msun, a dynamical timescale of 11 Myr, and a mass loss rate of $7.3 \times 10^{-4}$~\Msun~yr$^{-1}$. 
Particle acceleration at stellar winds have been discussed in, e.g., \cite{morlino_particle_2021,blasi_high-energy_2023}.
The maximum attainable energy depends on various conditions, 
and one example is provided as
\begin{equation}
    E_{\mathrm{max}} = 90 \left(\frac{L_c}{2 \, \mathrm{pc}}\right)^{-1} 
\left( \frac{\eta_B}{0.1} \right)^{\frac{1}{2}} 
\left( \frac{\dot{M}}{ 10^{-4}~M_\odot ~{\rm yr}^{-1} } \right)^{\frac{1}{5}} 
\left( \frac{v}{2~\kms} \right)^{ \frac{2}{5}}
\left( \frac{n}{1000~\cc} \right)^{ \frac{3}{10}} 
\left( \frac{t}{1~{\rm Myr}} \right)^{-\frac{2}{5}} 
~ \mathrm{TeV} , 
\label{eq:pmax}
\end{equation}
adjusted from Equation 18 in \cite{blasi_high-energy_2023}. 
This assumes that the scattering occurs in the inertial range of a Kraichnan-like turbulence. 
$\eta_B$ is a fractional ratio of the wind's kinetic energy transferred to the magnetic field.
Adopting the obtained parameters of Cloud E and $\eta_B = 0.1$, Equation~\ref{eq:pmax} yields the maximum energy of $\sim$90 TeV.
Although the mass loss rate of Cloud E and the tail-like structure ($\sim 10^{-4}$--$10^{-3}$~\Msun~yr$^{-1}$) is relatively high, the velocity is as low as 2~\kms, resulting in the low maximum energy.
It should be noted that under the different conditions (e.g., different turbulences such as Equations 13 and 16 in \cite{morlino_particle_2021}, which $E_{\rm max}$ is more dependent on the wind velocity),
the estimated maximum energy became much smaller than 90 TeV.
To conclude, it might be challenging to the AGB star (Cloud E) to accelerate particles beyond 100 TeV and to reconcile with the detection of $>100$~TeV gamma rays at 1LHAASO J0343$+$5254u with KM2A.
Nevertheless, it would be worth to theoretically and observationally investigate the possibility of efficient particle acceleration at AGB stars, which may emerge as a new class of CR accelerators.

As the new extended X-ray source has been detected in 1LHAASO J0343$+$5254u (WCDA) \citep{dikerby2024},
follow-up X-ray observations are needed for the other gamma-ray regions, 1LHAASO J0343$+$5254u (KM2A) and 1LHAASO J0339$+$5307 (KM2A), to explore and understand the nature of the J0341 complex.
High-resolution and wide-field-of-view X-ray observations, such as with AXIS\footnote{\url{https://blog.umd.edu/axis/}} \citep{reynolds_overview_2023,safi-harb_stellar_2023}, would be useful.
Further gamma-ray observations using Imaging Atmospheric Cherenkov Telescopes (IACTs) with better angular resolution will also help reveal a detailed gamma-ray structure and identify the accelerator. 
If the hadronic scenario is established, accompanying neutrinos can be tested by searching for them with next-generation observatories \citep[e.g., ][]{aiello_sensitivity_2019,clark_icecube-gen2_2021}.

\section{Conclusions}   \label{sec:conclusions}

LHAASO J0341$+$5258, discovered by LHAASO, is an unidentified, extended source and detected even at $>$100~TeV, making it one of the Galactic PeVatron candidate sources.
In order to reveal its nature, we have conducted molecular line observations by the NRO 45-m radio telescope.
Within the gamma-ray emitting region, we detected five sources,  Clouds A--E.
Clouds A--D are relatively nearby ($d\lesssim 1$ kpc), compact (the radii of 0.5--1.6 pc), light ($M \sim 5$--300 \Msun), and dense ($n \sim 200$--800 \cc).
Cloud E, a compact emission with optical and infrared counterparts IRAS 03392$+$5239, likely represents a CO envelope of an AGB star or a compact cloud clump.
We demonstrated that Cloud E might not be able to accelerate particles up to the PeV range.
We also analyzed the archival CGPS radio data.
The \HI\ clouds are extended and have comparable or lower mass than the CO clouds.
If the CO and \HI\ data are combined, the total proton number density is 700--5000~\cc.
In the 408 and 1420 MHz bands, there are only point-like sources and no apparent counterparts of the gamma rays in our RoI, and we derived flux upper limits of $\sim 10^{-13}$~\eflux.
If the gamma rays are of hadronic origin (i.e., associated with the detected molecular clouds), our findings indicate that the total proton budget is $\sim 10^{45}$ erg.
This is relatively small, suggesting that the gamma-ray emission might originate from molecular clouds illuminated by escaping CRs (CR-cloud interaction), rather than accelerators (such as SNR-cloud interaction).
Combined with follow-up X-ray observations \citep[e.g., ][]{dikerby2024} and gamma-ray observations with IACTs,
the origin of the gamma-ray radiation will be unveiled.


\begin{acknowledgments}
We are grateful to the anonymous referee for the helpful comments.
We thank Hidetoshi Sano, Tsunefumi Mizuno, and Laura Olivera-Nieto for the fruitful discussion.
We also thank Fiorenza Donato, Charles Hailey, Mattia Di Mauro, and Siliva Manconi for preparation of the proposals.
We are grateful to the staff of the Nobeyama Radio Observatory (NRO) for their outstanding support during the 45-m telescope observations. The NRO is a branch of the National Astronomical Observatory of Japan (NAOJ), National Institutes of Natural Sciences.
This work was supported by the Japan Society for the Promotion of Science (JSPS) KAKENHI grant Nos. JP22K14064 and JP24H01819 (N.T.).
This research has been partly supported by a grant from the Research Institute for Integrated Science, Kanagawa University (RIIS202403).
Support for K.M. and J.W. was provided by NASA through the XMMNC22 grant.
S.S.H. acknowledges support by the Natural Sciences and Engineering Research Council of Canada (NSERC) through the Canada Research Chairs and the Discovery Grants programs, and by the Canadian Space Agency.

\end{acknowledgments}

%

\vspace{5mm}
\facilities{NRO}



\clearpage

\appendix 

\section{Supplementary figures}
\label{sec:app}
\begin{figure*}[ht!]
 \begin{center}
  \includegraphics[width=\linewidth]{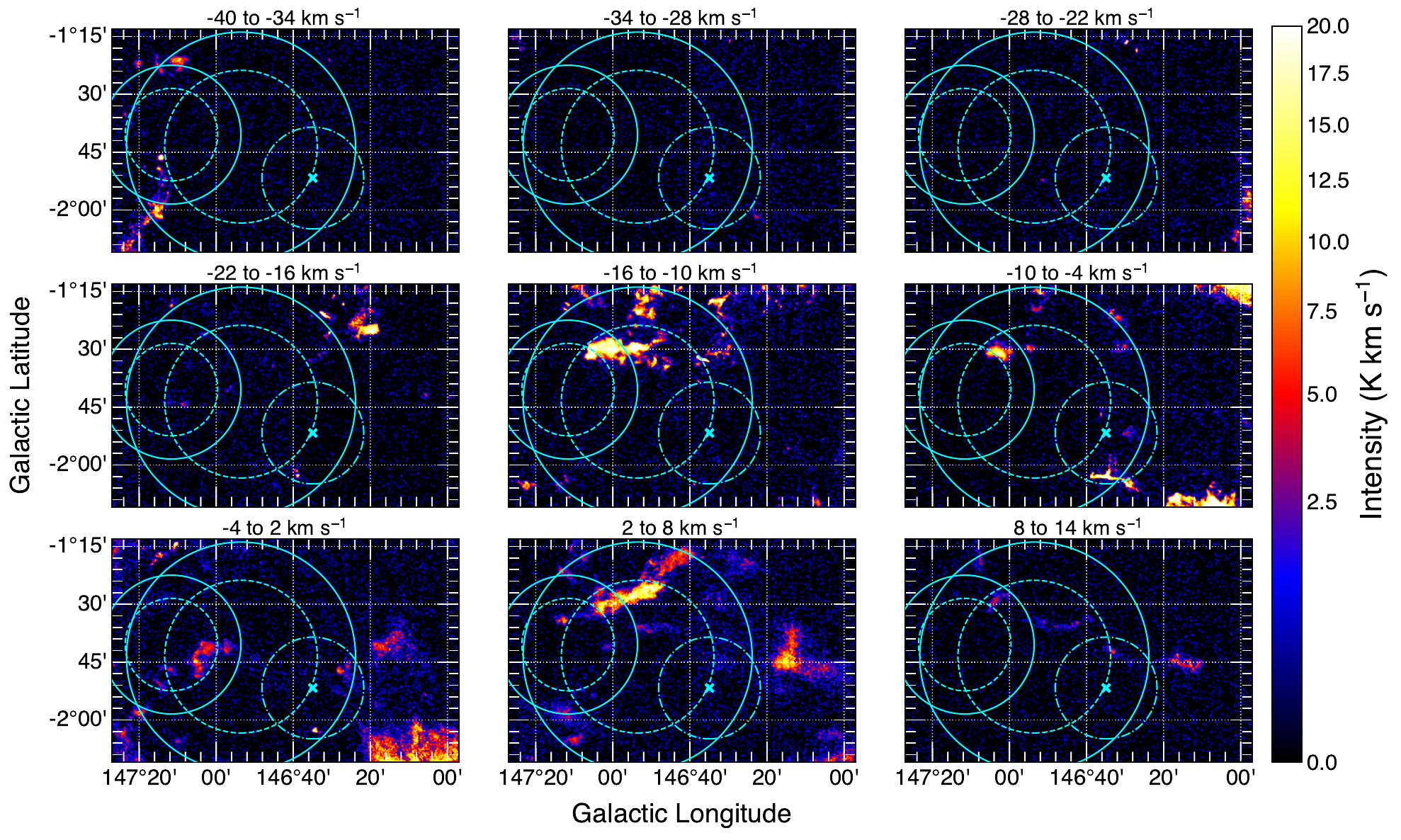}
 \end{center}
\caption{
\COa\ maps at different velocities with the interval of 6 \kms\ from $-$40 to 14 \kms. 
The dashed and solid circles indicate the radii of 39\% and 68\% containment of 1LHAASO J0343$+$5254u, respectively \citep{cao_first_2024}.
For 1LHAASO J0339$+$5307, the position and 95\% upper limit of the radius are shown with the cyan cross and dash-dotted circle, respectively. } 
\label{fig:grid}
\end{figure*}

\begin{figure}[ht!]
 \begin{center}
  \includegraphics[width=0.8\linewidth]{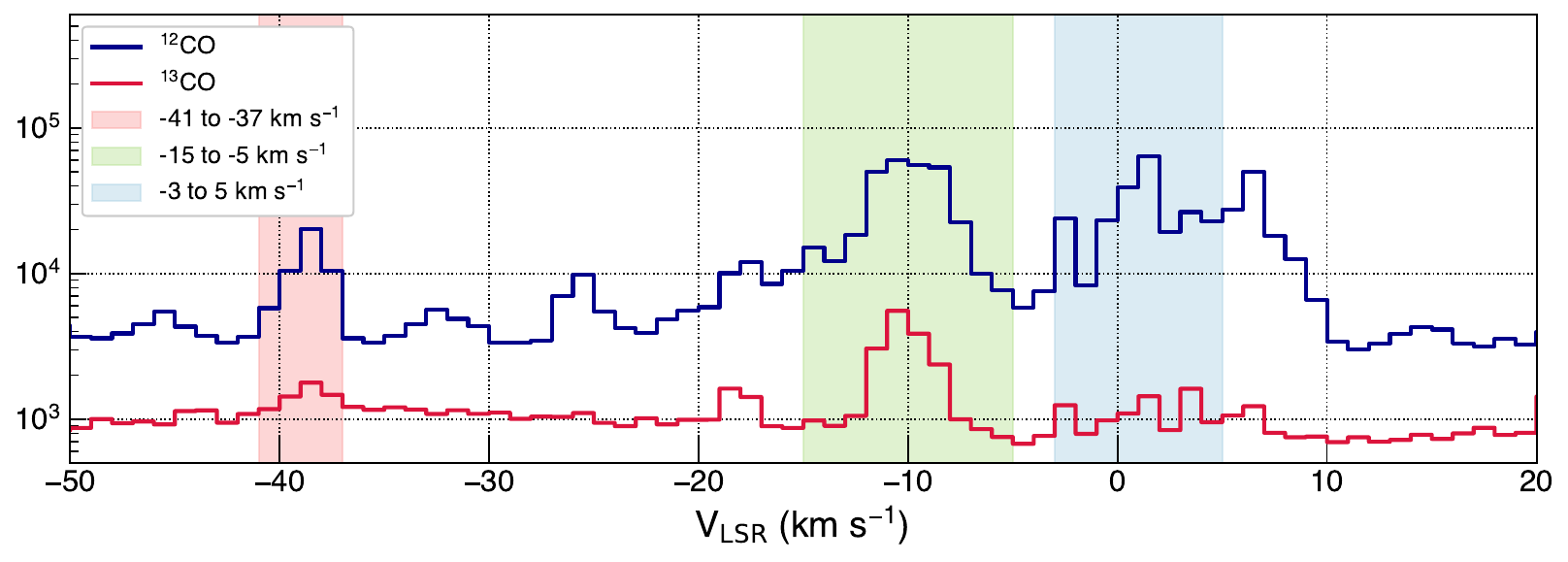}
 \end{center}
\caption{\COa\ and \COb\ velocity distributions extracted from the entire RoI.
}
\label{fig:spectrum}
\end{figure}

We show \COa\ maps at different velocity ranges in \figref{fig:grid} and the velocity spectrum from the entire RoI in \figref{fig:spectrum}.

\section{Molecular clouds outside the gamma-ray region}
\label{sec:outside}

Here, we present molecular clouds detected outside the LHAASO's gamma-ray extent.
It should be noted that the gamma-ray extension in this paper (\tabref{tab:overview}) indicates the 39\% containment radius assuming the two-dimensional Gaussian model (see \figref{fig:grid} for the 68\% containment radius).
Therefore, the clouds outside this radius might also be associated with the gamma rays.

\begin{figure}[ht!]
    \centering
    \includegraphics[width=0.75\linewidth]{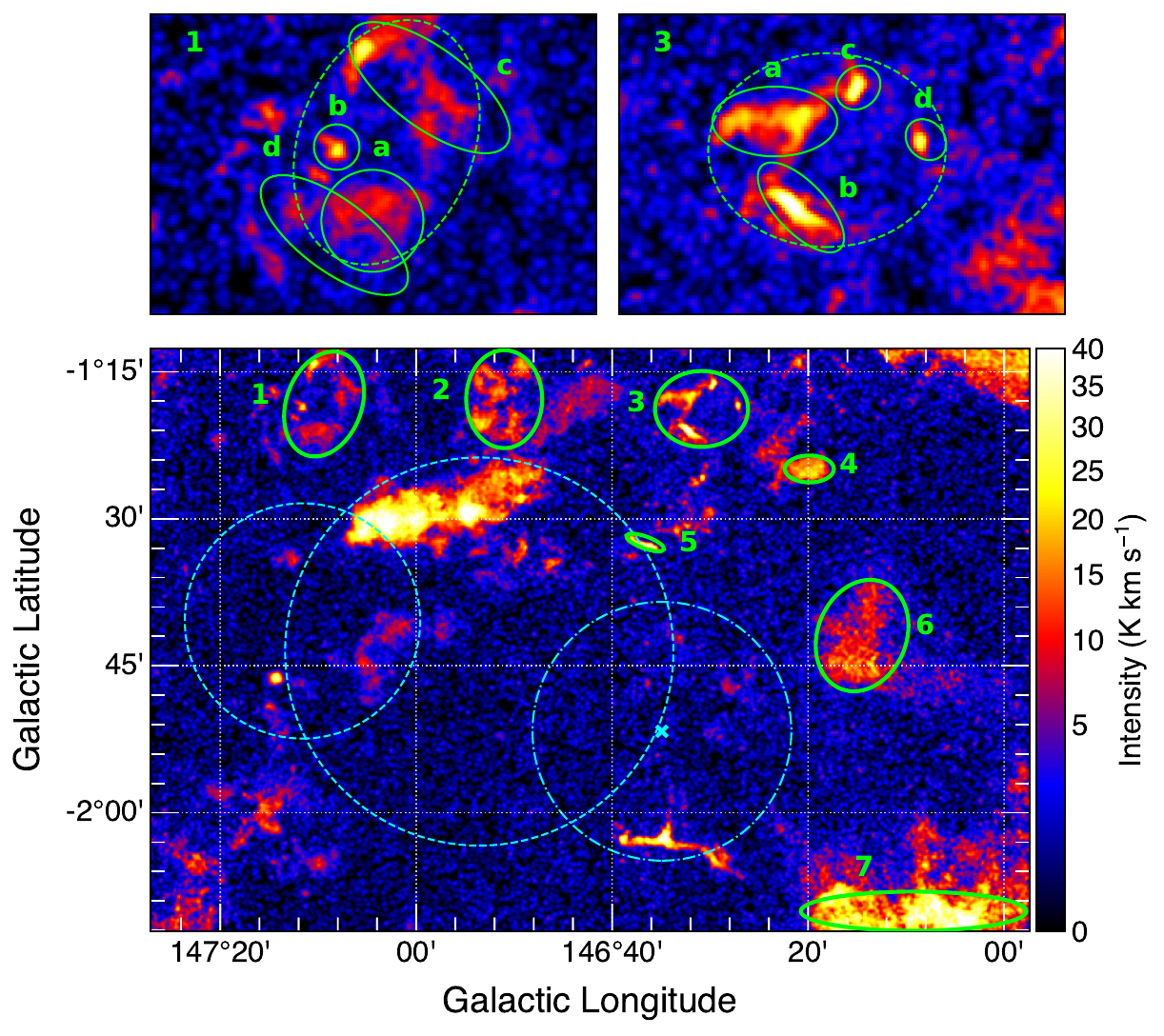}
    \caption{
    \COa\ map integrated over $-$40 to 10 \kms, shown with labels of the clouds outside the gamma-ray extent.
    The upper left and right panels illustrate the zoom-in images of Clouds 1 and 3, respectively.
    }
    \label{fig:outside}
\end{figure}

\begin{table*}[ht!]
\begin{center}
\scriptsize
  \caption{Physical parameters of clouds outside the J0341 region}
  \begin{tabular}{ccccccccccccc}
      \hline
 Cloud &     $\ell$ &    $b$ &  $V_{\rm LSR}$ &  $\Delta V_{\rm LSR}$ &  $d$ &  $R_{\rm min} ^*$ & $R_{\rm maj} ^*$ & $T_{\rm ex}$ &  $N(\mathrm{H}_2)$ &     $M(\mathrm{H}_2)$ &    $n(\mathrm{H}_2) ^\dagger$  \\ 
  &     (deg) &    (deg) &  (\kms) &  (\kms) &  (kpc) &  (pc) & (pc) &  (K) & (10$^{21}$ \columnd) &     (\Msun) &  (\cc)  \\ 
\hline
1a & 147.2 & -1.36 & -37.98 $\pm$ 0.00 & 1.07 $\pm$ 0.05 & 3.7 $\pm$ 0.9 & 2.4 & 2.4 & 5.1 $\pm$ 3.9 & 0.37 $\pm$ 0.18 & 150 $\pm$ 83 & 104 $\pm$ 59  \\ 
1b & 147.2 & -1.31 & -15.69 $\pm$ 0.18 & 4.35 $\pm$ 0.33 & 1.5 $\pm$ 0.8 & 0.37 & 0.37 & 4.1 $\pm$ 3.9 & 0.98 $\pm$ 0.18 & 9.4 $\pm$ 5.4 & 1800 $\pm$ 1100  \\ 
1c & 147.1 & -1.27 & -14.98 $\pm$ 0.00 & 1.23 $\pm$ 0.04 & 1.2 $\pm$ 0.5 & 0.68 & 1.7 & 4.3 $\pm$ 3.8 & 0.48 $\pm$ 0.18 & 38 $\pm$ 22 & 1200 $\pm$ 660  \\ 
1d & 147.2 & -1.37 & 7.47 $\pm$ 0.02 & 0.99 $\pm$ 0.02 & $<$0.2 & 0.077 & 0.19 & 4.4 $\pm$ 3.9 & 0.28 $\pm$ 0.18 & 0.29 $\pm$ 0.19 & 6200 $\pm$ 4100  \\ 
2 & 146.9 & -1.30 & -10.95 $\pm$ 0.03 & 1.97 $\pm$ 0.09 & 0.73 $\pm$ 0.6 & 1 & 1.3 & 5.4 $\pm$ 4.2 & 1.1 $\pm$ 0.24 & 100 $\pm$ 52 & 960$\pm$ 500  \\ 
3a & 146.6 & -1.30 & -11.34 $\pm$ 0.01 & 1.46 $\pm$ 0.01 & 0.87 $\pm$ 0.5 & 0.39 & 0.71 & 8.4 $\pm$ 4.1 & 1.6 $\pm$ 0.22 & 32 $\pm$ 18 & 5100 $\pm$ 2800  \\ 
3b & 146.5 & -1.35 & -17.06 $\pm$ 0.07 & 3.72 $\pm$ 0.19 & 1.5 $\pm$ 0.7 & 0.43 & 0.98 & 5.1 $\pm$ 3.9 & 1.5 $\pm$ 0.19 & 45 $\pm$ 20 & 5400 $\pm$ 2400  \\ 
3c & 146.5 & -1.27 & -22.70 $\pm$ 0.20 & 3.37 $\pm$ 0.50 & 2.1 $\pm$ 0.8 & 0.49 & 0.57 & 4.8 $\pm$ 3.9 & 1.2 $\pm$ 0.19 & 24 $\pm$ 9 & 2000 $\pm$ 780  \\ 
3d & 146.5 & -1.31 & -22.22 $\pm$ 0.14 & 3.86 $\pm$ 0.38 & 1.9 $\pm$ 0.7 & 0.44 & 0.54 & 4.4 $\pm$ 4.1 & 1.1 $\pm$ 0.21 & 18 $\pm$ 7.2 & 2000 $\pm$ 800  \\ 
4 & 146.3 & -1.42 & -17.91 $\pm$ 0.01 & 1.82 $\pm$ 0.04 &  1.4 $\pm$ 0.7 & 0.57 & 1 & 9 $\pm$ 4.2 & 2.5 $\pm$ 0.24 & 110 $\pm$ 48 & 5500 $\pm$ 2500  \\ 
5 & 146.6 & -1.54 & -13.73 $\pm$ 0.02 & 2.88 $\pm$ 0.05 & 1.1 $\pm$ 0.7 & 0.23 & 0.75 & 7.3 $\pm$ 4.3 & 2.4 $\pm$ 0.26 & 30 $\pm$ 14 & 25000 $\pm$ 12000  \\ 
6 & 146.2 & -1.70 & 6.24 $\pm$ 0.33 & 0.93 $\pm$ 0.65 &  $<$0.24 & 0.32 & 0.41 & 6.5 $\pm$ 4.3 & 0.97 $\pm$ 0.26 & 8.8 $\pm$ 2.4 & 2700 $\pm$ 730  \\ 
7 & 146.2 & -2.17 & -8.81 $\pm$ 0.02 & 1.76 $\pm$ 0.05 & 0.69 $\pm$ 0.4 & 0.37 & 2.1 & 8.5 $\pm$ 4.3 & 1.9 $\pm$ 0.26 & 110 $\pm$ 78 & 20000 $\pm$ 15000  \\ 
\hline
    \end{tabular}
    \label{tab:outside}
\end{center}    
\tablecomments{ \\
$^*$ $R_{\rm min}$ and $R_{\rm maj}$ indicate minor-axis and major-axis radii, respectively. \\
$\dag$ $n$ is derived by assuming a sphere and using the minor-axis radius. \\
}
\end{table*}

\bibliography{references,references2}{}

\begin{thebibliography}{}
\expandafter\ifx\csname natexlab\endcsname\relax\def\natexlab#1{#1}\fi
\providecommand{\url}[1]{\href{#1}{#1}}
\providecommand{\dodoi}[1]{doi:~\href{http://doi.org/#1}{\nolinkurl{#1}}}
\providecommand{\doeprint}[1]{\href{http://ascl.net/#1}{\nolinkurl{http://ascl.net/#1}}}
\providecommand{\doarXiv}[1]{\href{https://arxiv.org/abs/#1}{\nolinkurl{https://arxiv.org/abs/#1}}}

\bibitem[{Abdollahi {et~al.}(2020)Abdollahi, Acero, Ackermann, Ajello, Atwood, Axelsson, Baldini, Ballet, Barbiellini, Bastieri, Gonzalez, Bellazzini, Berretta, Bissaldi, Blandford, Bloom, Bonino, Bottacini, Brandt, Bregeon, Bruel, Buehler, Burnett, Buson, Cameron, Caputo, Caraveo, Casandjian, Castro, Cavazzuti, Charles, Chaty, Chen, Cheung, Chiaro, Ciprini, Cohen-Tanugi, Cominsky, Coronado-Blázquez, Costantin, Cuoco, Cutini, D'Ammando, DeKlotz, Luque, Palma, Desai, Digel, Lalla, Mauro, Venere, Domínguez, Dumora, Dirirsa, Fegan, Ferrara, Franckowiak, Fukazawa, Funk, Fusco, Gargano, Gasparrini, Giglietto, Giommi, Giordano, Giroletti, Glanzman, Green, Grenier, Griffin, Grondin, Grove, Guiriec, Harding, Hayashi, Hays, Hewitt, Horan, Jóhannesson, Johnson, Kamae, Kerr, Kocevski, Kovac'evic', Kuss, Landriu, Larsson, Latronico, Lemoine-Goumard, Li, Liodakis, Longo, Loparco, Lott, Lovellette, Lubrano, Madejski, Maldera, Malyshev, Manfreda, Marchesini, Marcotulli, Martí-Devesa, Martin, Massaro, Mazziotta, McEnery,
  Mereu, Meyer, Michelson, Mirabal, Mizuno, Monzani, Morselli, Moskalenko, Negro, Nuss, Ojha, Omodei, Orienti, Orlando, Ormes, Palatiello, Paliya, Paneque, Pei, Peña-Herazo, Perkins, Persic, Pesce-Rollins, Petrosian, Petrov, Piron, Poon, Porter, Principe, Rainò, Rando, Razzano, Razzaque, Reimer, Reimer, Remy, Reposeur, Romani, Parkinson, Schinzel, Serini, Sgrò, Siskind, Smith, Spandre, Spinelli, Strong, Suson, Tajima, Takahashi, Tak, Thayer, Thompson, Tibaldo, Torres, Torresi, Valverde, Klaveren, Zyl, Wood, Yassine, \& Zaharijas}]{4fgl}
Abdollahi, S., Acero, F., Ackermann, M., {et~al.} 2020, The Astrophysical Journal Supplement Series, 247, 33, \dodoi{10.3847/1538-4365/ab6bcb}

\bibitem[{Abe {et~al.}(2024)Abe, Abhir, Abhishek, Acero, Acharyya, Adam, Aguasca-Cabot, Agudo, Aguirre-Santaella, Alfaro, Alvarez-Crespo, Alves~Batista, Amans, Amato, Ambrosi, Ambrosino, Angüner, Aramo, Arcaro, Arrabito, Asano, Ascasíbar, Aschersleben, Augusto~Stuani, Backes, Balazs, Balbo, Ballet, Baquero~Larriva, Barbosa~Martins, Barres~de Almeida, Barrio, Batković, Batzofin, Baxter, Becerra~González, Beck, Beiske, Belmont, Benbow, Bernardini, Bernete, Bernlöhr, Berti, Bertucci, Beshley, Bhattacharjee, Bhattacharyya, Bi, Biederbeck, Biland, Bissaldi, Biteau, Blanch, Blazek, Bocchino, Boisson, Bolmont, Bonneau~Arbeletche, Bonnoli, Bonollo, Bordas, Bosnjak, Bottacini, Braiding, Bronzini, Brose, Brown, Brun, Brunelli, Bucciantini, Bulgarelli, Burelli, Burmistrov, Burton, Burtovoi, Bylund, Calisse, Campoy-Ordaz, Cantlay, Caproni, Capuzzo-Dolcetta, Caraveo, Caroff, Carosi, Carosi, Carquin, Carrasco, Cascone, Cassol, Castrejon, Castro-Tirado, Cerasole, Cerruti, Chadwick, Chambery, Chaty, Chen, Chernyakova,
  Chiavassa, Chytka, Cifuentes, Coimbra~Araujo, Conforti, Conte, Contreras, Cortina, Costa, Costantini, Cotter, Crestan, Cristofari, Cuevas, Curtis-Ginsberg, D'Aì, D'Amico, D'Ammando, Dadina, Dalchenko, David, Dazzi, de~Bony~de Lavergne, De~Caprio, De~Frondat~Laadim, de~Gouveia Dal~Pino, De~Lotto, De~Lucia, de~Martino, de~Menezes, de~Naurois, de~Ona~Wilhelmi, de~Souza, del Peral, Delgado~Giler, Delgado, Dell'aiera, Della~Valle, della Volpe, Depaoli, Di~Girolamo, Di~Piano, Di~Pierro, Di~Tria, Di~Venere, Díaz, Diebold, Dinesh, Djannati-Ataï, Djuvsland, Domínguez, Dominik, Donini, Dörner, Doro, dos Anjos, Dournaux, Duangchan, Dubos, Dubus, Duffy, Dumora, Dwarkadas, Ebr, Eckner, Egberts, Einecke, Elsässer, Emery, Errando, Escanuela, Escarate, Escobar~Godoy, Escudero, Esposito, Evoli, Falceta-Goncalves, Fattorini, Fegan, Feijen, Feng, Ferrand, Ferrarotto, Fiandrini, Fiasson, Filipovic, Fioretti, Fiori, Flores, Foffano, Font~Guiteras, Fontaine, Fröse, Fukazawa, Fukui, Funk, Furniss, Gaggero, Galanti, Galaz,
  \& Gallant}]{abe_prospects_2024}
Abe, S., Abhir, J., Abhishek, A., {et~al.} 2024, Journal of Cosmology and Astroparticle Physics, 2024, 081, \dodoi{10.1088/1475-7516/2024/10/081}

\bibitem[{Abeysekara {et~al.}(2020)Abeysekara, Archer, Benbow, Bird, Brose, Buchovecky, Buckley, Chromey, Cui, Daniel, Das, Dwarkadas, Falcone, Feng, Finley, Fortson, Gent, Gillanders, Giuri, Gueta, Hanna, Hassan, Hervet, Holder, Hughes, Humensky, Kaaret, Kar, Kelley-Hoskins, Kertzman, Kieda, Krause, Krennrich, Kumar, Lang, Maier, Moriarty, Mukherjee, Nievas-Rosillo, O'Brien, Ong, Park, Petrashyk, Pfrang, Pohl, Pueschel, Quinn, Ragan, Reynolds, Richards, Roache, Sadeh, Santander, Sembroski, Shahinyan, Sushch, Weinstein, Wilcox, Wilhelm, Williams, Williamson, Zitzer, \& Ghiotto}]{abeysekara_evidence_2020}
Abeysekara, A.~U., Archer, A., Benbow, W., {et~al.} 2020, The Astrophysical Journal, 894, 51, \dodoi{10.3847/1538-4357/ab8310}

\bibitem[{Ackermann {et~al.}(2013)Ackermann, Ajello, Allafort, Baldini, Ballet, Barbiellini, Baring, Bastieri, Bechtol, \& Bellazzini}]{Fermi2013_pi0}
Ackermann, M., Ajello, M., Allafort, A., {et~al.} 2013, Science, 339, 807, \dodoi{10.1126/science.1231160}

\bibitem[{Aharonian {et~al.}(2022)Aharonian, Ashkar, Backes, Barbosa~Martins, Becherini, Berge, Bi, Böttcher, de~Bony~de Lavergne, Bradascio, Brose, Brun, Bulik, Burger-Scheidlin, Cangemi, Caroff, Casanova, Cerruti, Chand, Chandra, Chen, Chibueze, Cristofari, Damascene~Mbarubucyeye, Djannati-Ataï, Ernenwein, Feijen, Fichet~de Clairfontaine, Fontaine, Funk, Gabici, Gallant, Ghafourizadeh, Giavitto, Giunti, Glawion, Glicenstein, Goswami, Grondin, Härer, Haupt, Hinton, Hörbe, Hofmann, Holch, Holler, Horns, Jamrozy, Joshi, Jung-Richardt, Kasai, Katarzyński, Katz, Khélifi, Kluźniak, Komin, Kosack, Kostunin, Kukec~Mezek, Lang, Le~Stum, Lemière, Lemoine-Goumard, Lenain, Leuschner, Lohse, Luashvili, Lypova, Mackey, Majumdar, Malyshev, Marandon, Marchegiani, Marcowith, Martí-Devesa, Marx, Maurin, Meyer, Mitchell, Moderski, Mohrmann, Montanari, Moulin, Muller, Murach, Nakashima, de~Naurois, Nayerhoda, Niemiec, Ohm, Olivera-Nieto, de~Ona~Wilhelmi, Ostrowski, Panny, Panter, Parsons, Peron, Prokhorov, Pühlhofer,
  Punch, Quirrenbach, Rauth, Reichherzer, Reimer, Reimer, Renaud, Reville, Rieger, Rowell, Rudak, Ruiz-Velasco, Sahakian, Salzmann, Sanchez, Santangelo, Sasaki, Schüssler, Schutte, Schwanke, Shapopi, Specovius, Spencer, Stawarz, Steenkamp, Steinmassl, Steppa, Sushch, Suzuki, Takahashi, Tanaka, Terrier, Thorpe-Morgan, Tsirou, Tsuji, Tuffs, Unbehaun, van Eldik, van Soelen, Vecchi, Veh, Venter, Vink, Wagner, White, Wierzcholska, Wong, Zacharias, Zargaryan, Zdziarski, Zhu, Zouari, Żywucka, Blackwell, Braiding, Burton, Cubuk, Filipović, Tothill, \& Wong}]{aharonian_deep_2022}
Aharonian, F., Ashkar, H., Backes, M., {et~al.} 2022, Astronomy and Astrophysics, 666, A124, \dodoi{10.1051/0004-6361/202244323}

\bibitem[{Aharonian(2013)}]{aharonian_gamma_2013}
Aharonian, F.~A. 2013, Astroparticle Physics, 43, 71, \dodoi{10.1016/j.astropartphys.2012.08.007}

\bibitem[{Aiello {et~al.}(2019)Aiello, Akrame, Ameli, G.~Anassontzis, Andre, Androulakis, Anghinolfi, Anton, Ardid, Aublin, Avgitas, Bagatelas, Barbarino, Baret, Barrios-Martí, Belias, Berbee, van~den Berg, Bertin, Biagi, Biagioni, Biernoth, Boumaaza, Bourret, Bouta, Bouwhuis, Bozza, Brânzaş, Bruchner, Bruijn, Brunner, Buis, Buompane, Busto, Calvo, Capone, Celli, Chabab, Chau, Cherubini, Chiarella, Chiarusi, Circella, Cocimano, Coelho, Coleiro, Molla, Coniglione, Coyle, Creusot, Cuttone, D’Onofrio, Dallier, De~Sio, Di~Palma, Díaz, Diego-Tortosa, Distefano, Domi, Donà, Donzaud, Dornic, Dörr, Durocher, Eberl, van Eijk, El~Bojaddaini, Eljarrari, Elsaesser, Enzenhöfer, Fermani, Ferrara, D.~Filipović, Fusco, Gal, Garcia, Garufi, Gialanella, Giorgio, Giuliante, Gozzini, Gracia, Graf, Grasso, Grégoire, Grella, Hallmann, Hamdaoui, van Haren, Heid, Heijboer, Hekalo, Hernández-Rey, Hofestädt, Illuminati, James, Jongen, de~Jong, de~Jong, Kadler, Kalaczyński, Kalekin, Katz, Khan~Chowdhury, Kießling,
  Koffeman, Kooijman, Kouchner, Kreter, Kulikovskiy, Kunhikannan~Kannichankandy, Lahmann, Larosa, Le~Breton, Leone, Leonora, Levi, Lincetto, Lonardo, Longhitano, Lopez~Coto, Lotze, Maderer, Maggi, Mańczak, Mannheim, Margiotta, Marinelli, Markou, Martin, Martínez-Mora, Martini, Marzaioli, Mele, Melis, Migliozzi, Migneco, Mijakowski, Miranda, Mollo, Morganti, Moser, Moussa, Muller, Musumeci, Nauta, Navas, Nicolau, Nielsen, Ó~Fearraigh, Organokov, Orlando, Ottonello, Panagopoulos, Papalashvili, Papaleo, Păvălaş, Pellegrino, Perrin-Terrin, Piattelli, Pikounis, Pisanti, Poiré, Polydefki, Popa, Post, Pradier, Pühlhofer, Pulvirenti, Quinn, Raffaelli, Randazzo, Razzaque, Real, Resvanis, Reubelt, Riccobene, Richer, Rigalleau, Rovelli, Saffer, Salvadori, Samtleben, Sánchez~Losa, Sanguineti, Santangelo, Santonocito, Sapienza, Schumann, Sciacca, Seneca, Sgura, Shanidze, Sharma, Simeone, Sinopoulou, Spisso, Spurio, Stavropoulos, Steijger, Stellacci, Strandberg, Stransky, Stüven, Taiuti, Tatone, Tayalati,
  Tenllado, Thakore, Trovato, Tzamariudaki, Tzanetatos, Van~Elewyck, Versari, Viola, Vivolo, Wilms, de~Wolf, Zaborov, Zornoza, \& Zúñiga}]{aiello_sensitivity_2019}
Aiello, S., Akrame, S.~E., Ameli, F., {et~al.} 2019, Astroparticle Physics, 111, 100, \dodoi{10.1016/j.astropartphys.2019.04.002}

\bibitem[{Alfaro {et~al.}(2024)Alfaro, Alvarez, Arteaga-Velázquez, Avila~Rojas, Ayala~Solares, Babu, Belmont-Moreno, Caballero-Mora, Capistrán, Carramiñana, Casanova, Cotti, Cotzomi, Coutiño~de León, De~la Fuente, Depaoli, Di~Lalla, Diaz~Hernandez, Dingus, DuVernois, Durocher, Díaz-Vélez, Engel, Espinoza, Fan, Fang, Fraija, Fraija, García-González, Garfias, Gonzalez~Muñoz, González, Goodman, Groetsch, Harding, Herzog, Hinton, Huang, Hueyotl-Zahuantitla, Hüntemeyer, Iriarte, Joshi, Kaufmann, Kieda, de~León, Lee, León~Vargas, Linnemann, Longinotti, Luis-Raya, Malone, Martinez, Martínez-Castro, Matthews, Miranda-Romagnoli, Morales-Soto, Moreno, Mostafá, Nayerhoda, Nellen, Newbold, Nisa, Noriega-Papaqui, Olivera-Nieto, Omodei, Osorio, Pérez~Araujo, Pérez-Pérez, Rho, Rosa-González, Ruiz-Velasco, Salazar, Salazar-Gallegos, Sandoval, Schneider, Serna-Franco, Smith, Son, Springer, Tibolla, Tollefson, Torres, Torres-Escobedo, Turner, Ureña-Mena, Varela, Villaseñor, Wang, Watson, Willox,
  Yun-Cárcamo, \& Zhou}]{alfaro_ultra-high-energy_2024}
Alfaro, R., Alvarez, C., Arteaga-Velázquez, J.~C., {et~al.} 2024, Nature, 634, 557, \dodoi{10.1038/s41586-024-07995-9}

\bibitem[{Amenomori {et~al.}(2005)Amenomori, Ayabe, Chen, Cui, Danzengluobu, Ding, Ding, Feng, Feng, Gao, Geng, Guo, He, He, Hibino, Hotta, Hu, Hu, Huang, Huang, Jia, Kajino, Kasahara, Katayose, Kato, Kawata, Labaciren, Le, Li, Lu, Lu, Meng, Mizutani, Mu, Munakata, Nagai, Nanjo, Nishizawa, Ohnishi, Ohta, Onuma, Ouchi, Ozawa, Ren, Saito, Sakata, Sasaki, Shibata, Shiomi, Shirai, Sugimoto, Takashima, Takita, Tan, Tateyama, Torii, Tsuchiya, Udo, Utsugi, Wang, Wang, Wang, Wu, Xue, Yamamoto, Yan, Yang, Yasue, Ye, Yu, Yuan, Yuda, Zhang, Zhang, Zhang, Zhang, Zhang, Zhang, Zhaxisangzhu, Zhou, \& Collaboration)}]{amenomori_northern_2005}
Amenomori, M., Ayabe, S., Chen, D., {et~al.} 2005, The Astrophysical Journal, 633, 1005, \dodoi{10.1086/491612}

\bibitem[{Arimoto {et~al.}(1996)Arimoto, Sofue, \& Tsujimoto}]{arimoto_co--h2_1996}
Arimoto, N., Sofue, Y., \& Tsujimoto, T. 1996, Publications of the Astronomical Society of Japan, 48, 275, \dodoi{10.1093/pasj/48.2.275}

\bibitem[{Bangale \& Wang(2023)}]{bangale_searching_2023}
Bangale, P., \& Wang, X. 2023, Searching for {TeV} emission from {LHAASO} {J0341}+5258 with {VERITAS} and {HAWC}, \dodoi{10.48550/arXiv.2308.15643}

\bibitem[{Blasi \& Morlino(2023)}]{blasi_high-energy_2023}
Blasi, P., \& Morlino, G. 2023, Monthly Notices of the Royal Astronomical Society, 523, 4015, \dodoi{10.1093/mnras/stad1662}

\bibitem[{Bolatto {et~al.}(2013)Bolatto, Wolfire, \& Leroy}]{bolatto_co--h2_2013}
Bolatto, A.~D., Wolfire, M., \& Leroy, A.~K. 2013, Annual Review of Astronomy and Astrophysics, 51, 207, \dodoi{10.1146/annurev-astro-082812-140944}

\bibitem[{Brand \& Blitz(1993)}]{brand_velocity_1993}
Brand, J., \& Blitz, L. 1993, Astronomy and Astrophysics, 275, 67.
\newblock \url{https://ui.adsabs.harvard.edu/abs/1993A&A...275...67B}

\bibitem[{Cao {et~al.}(2021{\natexlab{a}})Cao, Aharonian, An, {Axikegu}, Bai, Bai, Bao, Bastieri, Bi, Bi, Cai, Cai, Cao, Chang, Chang, Chang, Chen, Chen, Chen, Chen, Chen, Chen, Chen, Chen, Chen, Chen, Chen, Chen, Chen, Cheng, Cheng, Cui, Cui, Cui, Dai, Dai, Dai, {Danzengluobu}, della Volpe, D′Ettorre~Piazzoli, Dong, Fan, Fan, Fan, Fang, Fang, Feng, Feng, Feng, Feng, Gao, Gao, Gao, Gao, Ge, Geng, Gong, Gou, Gu, Guo, Guo, Guo, Guo, Han, He, He, He, He, He, He, Heller, Hor, Hou, Hou, Hu, Hu, Hu, Hu, Huang, Huang, Huang, Huang, Huang, Ji, Ji, Jia, Jiang, Jiang, Jin, Kuleshov, Levochkin, Li, Li, Li, Li, Li, Li, Li, Li, Li, Li, Li, Li, Li, Li, Li, Li, Li, Liang, Liang, Lin, Liu, Liu, Liu, Liu, Liu, Liu, Liu, Liu, Liu, Liu, Liu, Liu, Liu, Liu, Liu, Long, Lu, Lv, Ma, Ma, Ma, Mao, Masood, Mitthumsiri, Montaruli, Nan, Pang, Pattarakijwanich, Pei, Qi, Ruffolo, Rulev, Sáiz, Shao, Shchegolev, Sheng, Shi, Song, Stenkin, Stepanov, Sun, Sun, Sun, Tam, Tang, Tian, Wang, Wang, Wang, Wang, Wang, Wang, Wang, Wang, Wang,
  Wang, Wang, Wang, Wang, Wang, Wang, Wang, Wang, Wang, Wang, Wang, Wang, Wei, Wei, Wei, Wen, Wu, Wu, Wu, Wu, Wu, Xi, Xia, Xia, Xiang, Xiao, Xiao, Xin, Xin, Xing, Xu, Xu, Xue, Yan, Yang, Yang, Yang, Yang, Yang, Yang, Yang, Yao, Yao, Ye, Yin, Yin, You, You, Yu, Yuan, Zeng, Zeng, Zeng, Zeng, Zha, Zhai, Zhang, Zhang, Zhang, Zhang, Zhang, Zhang, Zhang, Zhang, Zhang, Zhang, Zhang, Zhang, Zhang, Zhang, Zhang, Zhang, Zhang, Zhang, Zhang, Zhao, Zhao, Zhao, Zhao, Zhao, Zheng, Zheng, Zhou, Zhou, Zhou, Zhou, Zhou, Zhou, Zhu, Zhu, Zhu, Zhu, \& Zuo}]{cao_ultrahigh-energy_2021}
Cao, Z., Aharonian, F.~A., An, Q., {et~al.} 2021{\natexlab{a}}, Nature, \dodoi{10.1038/s41586-021-03498-z}

\bibitem[{Cao {et~al.}(2021{\natexlab{b}})Cao, Aharonian, An, {Axikegu}, Bai, Bai, Bao, Bastieri, Bi, Bi, Cai, Cai, Cao, Chang, Chang, Chen, Chen, Chen, Chen, Chen, Chen, Chen, Chen, Chen, Chen, Chen, Chen, Chen, Chen, Cheng, Cheng, Cui, Cui, Cui, Piazzoli, Dai, Dai, Dai, {Danzengluobu}, Volpe, Dong, Duan, Fan, Fan, Fan, Fang, Fang, Feng, Feng, Feng, Feng, Gao, Gao, Gao, Gao, Gao, Ge, Geng, Gong, Gou, Gu, Guo, Guo, Guo, Guo, Guo, Han, He, He, He, He, He, He, Heller, Hor, Hou, Hu, Hu, Hu, Hu, Huang, Huang, Huang, Huang, Huang, Huang, Ji, Ji, Jia, Jiang, Jiang, Jin, Ke, Kuleshov, Levochkin, Li, Li, Li, Li, Li, Li, Li, Li, Li, Li, Li, Li, Li, Li, Li, Li, Li, Liang, Liang, Lin, Liu, Liu, Liu, Liu, Liu, Liu, Liu, Liu, Liu, Liu, Liu, Liu, Liu, Liu, Liu, Liu, Long, Lu, Lv, Ma, Ma, Ma, Mao, Masood, Min, Mitthumsiri, Montaruli, Nan, Pang, Pattarakijwanich, Pei, Qi, Qi, Qiao, Qin, Ruffolo, Rulev, Sáiz, Shao, Shchegolev, Sheng, Shi, Song, Stenkin, Stepanov, Su, Sun, Sun, Sun, Tam, Tang, Tian, Wang, Wang, Wang, Wang,
  Wang, Wang, Wang, Wang, Wang, Wang, Wang, Wang, Wang, Wang, Wang, Wang, Wang, Wang, Wang, Wang, Wang, Wang, Wei, Wei, Wei, Wen, Wu, Wu, Wu, Wu, Wu, Xi, Xia, Xia, Xiang, Xiao, Xiao, Xiao, Xin, Xin, Xing, Xu, Xu, Xue, Yan, Yan, Yang, Yang, Yang, Yang, Yang, Yang, Yang, Yao, Yao, Ye, Yin, Yin, You, You, Yu, Yuan, Zeng, Zeng, Zeng, Zeng, Zha, Zhai, Zhang, Zhang, Zhang, Zhang, Zhang, Zhang, Zhang, Zhang, Zhang, Zhang, Zhang, Zhang, Zhang, Zhang, Zhang, Zhang, Zhang, Zhang, Zhang, Zhao, Zhao, Zhao, Zhao, Zhao, Zheng, Zheng, Zhou, Zhou, Zhou, Zhou, Zhou, Zhou, Zhu, Zhu, Zhu, Zhu, \& Zuo}]{cao_discovery_2021}
Cao, Z., Aharonian, F., An, Q., {et~al.} 2021{\natexlab{b}}, The Astrophysical Journal Letters, 917, L4, \dodoi{10.3847/2041-8213/ac0fd5}

\bibitem[{Cao {et~al.}(2024{\natexlab{a}})Cao, Aharonian, {Axikegu}, Bai, Bao, Bastieri, Bi, Bi, Bian, Bukevich, Cao, Cao, Cao, Chang, Chang, Chen, Chen, Chen, Chen, Chen, Chen, Chen, Chen, Chen, Chen, Chen, Chen, Chen, Chen, Cheng, Cheng, Cui, Cui, Cui, Cui, Dai, Dai, Dai, {Danzengluobu}, Dong, Duan, Fan, Fan, Fang, Fang, Fang, Feng, Feng, Feng, Feng, Feng, Feng, Feng, Gabici, Gao, Gao, Gao, Gao, Gao, Ge, Geng, Giacinti, Gong, Gou, Gu, Guo, Guo, Guo, Guo, Han, Hasan, He, He, He, He, Hor, Hou, Hou, Hou, Hu, Hu, Hu, Huang, Huang, Huang, Huang, Huang, Huang, Ji, Jia, Jia, Jiang, Jiang, Jiang, Jin, Kang, Karpikov, Kuleshov, Kurinov, Li, Li, Li, Li, Li, Li, Li, Li, Li, Li, Li, Li, Li, Li, Li, Li, Li, Li, Li, Liang, Liang, Lin, Liu, Liu, Liu, Liu, Liu, Liu, Liu, Liu, Liu, Liu, Liu, Liu, Liu, Liu, Luo, Luo, Lv, Ma, Ma, Ma, Mao, Min, Mitthumsiri, Mu, Nan, Neronov, Ou, Pattarakijwanich, Pei, Qi, Qi, Qiao, Qin, Raza, Ruffolo, Sáiz, Saeed, Semikoz, Shao, Shchegolev, Sheng, Shu, Song, Stenkin, Stepanov, Su, Sun, Sun,
  Sun, Sun, Takata, Tam, Tang, Tang, Tang, Tian, Wang, Wang, Wang, Wang, Wang, Wang, Wang, Wang, Wang, Wang, Wang, Wang, Wang, Wang, Wang, Wang, Wang, Wang, Wang, Wang, Wang, Wang, \& Wei}]{cao_evidence_2024}
Cao, Z., Aharonian, F., {Axikegu}, {et~al.} 2024{\natexlab{a}}, Science Bulletin, 69, 2833, \dodoi{10.1016/j.scib.2024.07.017}

\bibitem[{Cao {et~al.}(2024{\natexlab{b}})Cao, Aharonian, An, {Axikegu}, Bai, Bao, Bastieri, Bi, Bi, Cai, Cao, Cao, Cao, Chang, Chang, Chen, Chen, Chen, Chen, Chen, Chen, Chen, Chen, Chen, Chen, Chen, Chen, Cheng, Cheng, Cui, Cui, Cui, Cui, Dai, Dai, Dai, {Danzengluobu}, Della~Volpe, Dong, Duan, Fan, Fan, Fang, Fang, Feng, Feng, Feng, Feng, Feng, Gabici, Gao, Gao, Gao, Gao, Gao, Gao, Ge, Geng, Giacinti, Gong, Gou, Gu, Guo, Guo, Guo, Guo, Han, He, He, He, He, He, Heller, Hor, Hou, Hou, Hou, Hu, Hu, Hu, Huang, Huang, Huang, Huang, Huang, Huang, Huang, Ji, Jia, Jia, Jiang, Jiang, Jiang, Jin, Kang, Ke, Kuleshov, Kurinov, Li, Li, Li, Li, Li, Li, Li, Li, Li, Li, Li, Li, Li, Li, Li, Li, Li, Li, Li, Liang, Liang, Lin, Liu, Liu, Liu, Liu, Liu, Liu, Liu, Liu, Liu, Liu, Liu, Liu, Liu, Liu, Lu, Luo, Lv, Ma, Ma, Ma, Mao, Min, Mitthumsiri, Mu, Nan, Neronov, Ou, Pang, Pattarakijwanich, Pei, Qi, Qi, Qiao, Qin, Ruffolo, Sáiz, Semikoz, Shao, Shao, Shchegolev, Sheng, Shu, Song, Stenkin, Stepanov, Su, Sun, Sun, Sun, Tam, Tang,
  Tang, Tian, Wang, Wang, Wang, Wang, Wang, Wang, Wang, Wang, Wang, Wang, Wang, Wang, Wang, Wang, Wang, Wang, Wang, Wang, Wang, Wang, Wang, Wei, Wei, Wei, Wen, Wu, \& Wu}]{cao_first_2024}
Cao, Z., Aharonian, F., An, Q., {et~al.} 2024{\natexlab{b}}, The Astrophysical Journal Supplement Series, 271, 25, \dodoi{10.3847/1538-4365/acfd29}

\bibitem[{Clark(2021)}]{clark_icecube-gen2_2021}
Clark, B. 2021, Journal of Instrumentation, 16, C10007, \dodoi{10.1088/1748-0221/16/10/C10007}

\bibitem[{Cox {et~al.}(2012)Cox, Kerschbaum, van Marle, Decin, Ladjal, Mayer, Groenewegen, van Eck, Royer, Ottensamer, Ueta, Jorissen, Mecina, Meliani, Luntzer, Blommaert, Posch, Vandenbussche, \& Waelkens}]{cox_far-infrared_2012}
Cox, N. L.~J., Kerschbaum, F., van Marle, A.~J., {et~al.} 2012, Astronomy and Astrophysics, 537, A35, \dodoi{10.1051/0004-6361/201117910}

\bibitem[{Dame {et~al.}(2001)Dame, Hartmann, \& Thaddeus}]{dame_milky_2001}
Dame, T.~M., Hartmann, D., \& Thaddeus, P. 2001, The Astrophysical Journal, 547, 792, \dodoi{10.1086/318388}

\bibitem[{Dame \& Thaddeus(2022)}]{dame_co_2022}
Dame, T.~M., \& Thaddeus, P. 2022, The Astrophysical Journal Supplement Series, 262, 5, \dodoi{10.3847/1538-4365/ac7e53}

\bibitem[{de~la Fuente {et~al.}(2023)de~la Fuente, Toledano-Juarez, Kawata, Trinidad, Tafoya, Sano, Tokuda, Nishimura, Onishi, Sako, Hona, Ohnishi, \& Takita}]{de_la_fuente_detection_2023}
de~la Fuente, E., Toledano-Juarez, I., Kawata, K., {et~al.} 2023, Publications of the Astronomical Society of Japan, 75, 546, \dodoi{10.1093/pasj/psad018}

\bibitem[{De~Sarkar \& Majumdar(2023)}]{de_sarkar_dissecting_2023}
De~Sarkar, A., \& Majumdar, P. 2023, Dissecting the emission from {LHAASO} {J0341}+5258: implications for future multi-wavelength observations, \dodoi{10.48550/arXiv.2309.04729}

\bibitem[{Dickey \& Lockman(1990)}]{dickey_h_1990}
Dickey, J.~M., \& Lockman, F.~J. 1990, Annual Review of Astronomy and Astrophysics, 28, 215, \dodoi{10.1146/annurev.aa.28.090190.001243}

\bibitem[{DiKerby {et~al.}(2025)DiKerby, Zhang, \& Mori}]{dikerby2024}
DiKerby, S., Zhang, S., \& Mori, K. 2025, ApJ, in prep.

\bibitem[{Ferrand \& Safi-Harb(2012)}]{SNRcat}
Ferrand, G., \& Safi-Harb, S. 2012, Advances in Space Research, 49, 1313, \dodoi{10.1016/j.asr.2012.02.004}

\bibitem[{Funk(2015)}]{Funk2015}
Funk, S. 2015, Annual Review of Nuclear and Particle Science, 65, 245, \dodoi{10.1146/annurev-nucl-102014-022036}

\bibitem[{Garay \& Lizano(1999)}]{garay_massive_1999}
Garay, G., \& Lizano, S. 1999, Publications of the Astronomical Society of the Pacific, 111, 1049, \dodoi{10.1086/316416}

\bibitem[{Giunti {et~al.}(2022)Giunti, Acero, Khélifi, Kosack, Lemière, \& Terrier}]{giunti_constraining_2022}
Giunti, L., Acero, F., Khélifi, B., {et~al.} 2022, Astronomy \& Astrophysics, 667, A130, \dodoi{10.1051/0004-6361/202244696}

\bibitem[{Grenier {et~al.}(2015)Grenier, Black, \& Strong}]{grenier_nine_2015}
Grenier, I.~A., Black, J.~H., \& Strong, A.~W. 2015, Annual Review of Astronomy and Astrophysics, 53, 199, \dodoi{10.1146/annurev-astro-082214-122457}

\bibitem[{Grenier {et~al.}(2005)Grenier, Casandjian, \& Terrier}]{grenier_unveiling_2005}
Grenier, I.~A., Casandjian, J.-M., \& Terrier, R. 2005, Science, 307, 1292, \dodoi{10.1126/science.1106924}

\bibitem[{{HAWC Collaboration} {et~al.}(2020){HAWC Collaboration}, Abeysekara, Albert, Alfaro, Angeles~Camacho, Arteaga-Velázquez, Arunbabu, Avila~Rojas, Ayala~Solares, Baghmanyan, Belmont-Moreno, BenZvi, Brisbois, Caballero-Mora, Capistrán, Carramiñana, Casanova, Cotti, Cotzomi, Coutiño~de León, De~la Fuente, de~León, Dichiara, Dingus, DuVernois, Díaz-Vélez, Ellsworth, Engel, Espinoza, Fleischhack, Fraija, Galván-Gámez, Garcia, García-González, Garfias, González, Goodman, Harding, Hernandez, Hinton, Hona, Huang, Hueyotl-Zahuantitla, Hüntemeyer, Iriarte, Jardin-Blicq, Joshi, Kaufmann, Kieda, Lara, Lee, León~Vargas, Linnemann, Longinotti, Luis-Raya, Lundeen, López-Coto, Malone, Marinelli, Martinez, Martinez-Castellanos, Martínez-Castro, Martínez-Huerta, Matthews, Miranda-Romagnoli, Morales-Soto, Moreno, Mostafá, Nayerhoda, Nellen, Newbold, Nisa, Noriega-Papaqui, Peisker, Pérez-Pérez, Pretz, Ren, Rho, Rivière, Rosa-González, Rosenberg, Ruiz-Velasco, Salesa~Greus, Sandoval, Schneider,
  Schoorlemmer, Sinnis, Smith, Springer, Surajbali, Tabachnick, Tanner, Tibolla, Tollefson, Torres, Torres-Escobedo, Villaseñor, Weisgarber, Wood, Yapici, Zhang, \& Zhou}]{hawc_collaboration_multiple_2020}
{HAWC Collaboration}, Abeysekara, A., Albert, A., {et~al.} 2020, Physical Review Letters, 124, 021102, \dodoi{10.1103/PhysRevLett.124.021102}

\bibitem[{{H.E.S.S. COLLABORATION}(2024)}]{hess_collaboration_acceleration_2024}
{H.E.S.S. COLLABORATION}. 2024, Science, 383, 402, \dodoi{10.1126/science.adi2048}

\bibitem[{{H.E.S.S. Collaboration} {et~al.}(2018){H.E.S.S. Collaboration}, Abdalla, Abramowski, Aharonian, Ait~Benkhali, Angüner, Arakawa, Arrieta, Aubert, Backes, Balzer, Barnard, Becherini, Becker~Tjus, Berge, Bernhard, Bernlöhr, Blackwell, Böttcher, Boisson, Bolmont, Bonnefoy, Bordas, Bregeon, Brun, Brun, Bryan, Büchele, Bulik, Capasso, Caroff, Carosi, Casanova, Cerruti, Chakraborty, Chaves, Chen, Chevalier, Colafrancesco, Condon, Conrad, Davids, Decock, Deil, Devin, deWilt, Dirson, Djannati-Ataï, Donath, Drury, Dutson, Dyks, Edwards, Egberts, Emery, Ernenwein, Eschbach, Farnier, Fegan, Fernandes, Fernandez, Fiasson, Fontaine, Funk, Füßling, Gabici, Gallant, Garrigoux, Gaté, Giavitto, Giebels, Glawion, Glicenstein, Gottschall, Grondin, Hahn, Haupt, Hawkes, Heinzelmann, Henri, Hermann, Hinton, Hofmann, Hoischen, Holch, Holler, Horns, Ivascenko, Iwasaki, Jacholkowska, Jamrozy, Jankowsky, Jankowsky, Jingo, Jouvin, Jung-Richardt, Kastendieck, Katarzyński, Katsuragawa, Katz, Kerszberg, Khangulyan,
  Khélifi, King, Klepser, Klochkov, Kluźniak, Komin, Kosack, Krakau, Kraus, Krüger, Laffon, Lamanna, Lau, Lees, Lefaucheur, Lemière, Lemoine-Goumard, Lenain, Leser, Lohse, Lorentz, Liu, López-Coto, Lypova, Malyshev, Marandon, Marcowith, Mariaud, Marx, Maurin, Maxted, Mayer, Meintjes, Meyer, Mitchell, Moderski, Mohamed, Mohrmann, Morå, Moulin, Murach, Nakashima, de~Naurois, Ndiyavala, Niederwanger, Niemiec, Oakes, O’Brien, Odaka, Ohm, Ostrowski, Oya, Padovani, Panter, Parsons, Pekeur, Pelletier, Perennes, Petrucci, Peyaud, Piel, Pita, Poireau, Poon, Prokhorov, Prokoph, Pühlhofer, Punch, Quirrenbach, Raab, Rauth, Reimer, Reimer, Renaud, de~los Reyes, Rieger, Rinchiuso, Romoli, Rowell, Rudak, Rulten, Safi-Harb, Sahakian, Saito, Sanchez, Santangelo, Sasaki, Schlickeiser, Schüssler, Schulz, Schwanke, Schwemmer, Seglar-Arroyo, Settimo, Seyffert, Shafi, Shilon, Shiningayamwe, Simoni, Sol, Spanier, Spir-Jacob, Stawarz, Steenkamp, Stegmann, Steppa, Sushch, Takahashi, Tavernet, Tavernier, Taylor, Terrier,
  Tibaldo, Tiziani, Tluczykont, Trichard, Tsirou, Tsuji, Tuffs, Uchiyama, van~der Walt, van Eldik, van Rensburg, van Soelen, Vasileiadis, Veh, Venter, Viana, Vincent, Vink, Voisin, Völk, Vuillaume, Wadiasingh, Wagner, Wagner, Wagner, White, Wierzcholska, Willmann, Wörnlein, Wouters, Yang, Zaborov, Zacharias, Zanin, Zdziarski, Zech, Zefi, Ziegler, Zorn, \& Żywucka}]{HESS2018_SNR}
{H.E.S.S. Collaboration}, Abdalla, H., Abramowski, A., {et~al.} 2018, Astronomy \& Astrophysics, 612, A3, \dodoi{10.1051/0004-6361/201732125}

\bibitem[{Kafexhiu {et~al.}(2014)Kafexhiu, Aharonian, Taylor, \& Vila}]{Kafexhiu2014}
Kafexhiu, E., Aharonian, F., Taylor, A.~M., \& Vila, G.~S. 2014, 90, 123014, \dodoi{10.1103/PhysRevD.90.123014}

\bibitem[{Kar \& Gupta(2022)}]{kar_ultrahigh-energy_2022}
Kar, A., \& Gupta, N. 2022, The Astrophysical Journal, 926, 110, \dodoi{10.3847/1538-4357/ac4757}

\bibitem[{Kuno {et~al.}(2011)Kuno, Takano, Iono, Nakajima, Iwashita, Handa, Hatsukade, Higuchi, Hirota, Ishikawa, Kaneko, Kawaguchi, Kawabe, Kimura, Kohno, Maekawa, Mikoshiba, Miyazawa, Miyazawa, Muraoka, Ogawa, Onodera, Saito, Takahashi, \& Yonezu}]{kuno_new_2011}
Kuno, N., Takano, S., Iono, D., {et~al.} 2011, in 2011 {XXXth} {URSI} {General} {Assembly} and {Scientific} {Symposium}, 1--4, \dodoi{10.1109/URSIGASS.2011.6051296}

\bibitem[{Kutner \& Ulich(1981)}]{kutner_recommendations_1981}
Kutner, M.~L., \& Ulich, B.~L. 1981, The Astrophysical Journal, 250, 341, \dodoi{10.1086/159380}

\bibitem[{{LHAASO Collaboration}(2010)}]{lhaaso_collaboration_future_2010}
{LHAASO Collaboration}. 2010, Chinese Physics C, 34, 249, \dodoi{10.1088/1674-1137/34/2/018}

\bibitem[{{LHAASO Collaboration}(2024{\natexlab{a}})}]{lhaaso_2024_cygnus}
---. 2024{\natexlab{a}}, Science Bulletin, 69, 449, \dodoi{10.1016/j.scib.2023.12.040}

\bibitem[{{LHAASO Collaboration}(2024{\natexlab{b}})}]{lhaaso_2024_microquasars}
---. 2024{\natexlab{b}}, Ultrahigh-{Energy} {Gamma}-ray {Emission} {Associated} with {Black} {Hole}-{Jet} {Systems}, \dodoi{10.48550/arXiv.2410.08988}

\bibitem[{McKee \& Ostriker(2007)}]{mckee_theory_2007}
McKee, C.~F., \& Ostriker, E.~C. 2007, Annual Review of Astronomy and Astrophysics, 45, 565, \dodoi{10.1146/annurev.astro.45.051806.110602}

\bibitem[{Minamidani {et~al.}(2016)Minamidani, Nishimura, Miyamoto, Kaneko, Iwashita, Miyazawa, Nishitani, Wada, Fujii, Takahashi, Iizuka, Ogawa, Kimura, Kozuki, Hasegawa, Matsuo, Fujita, Ohashi, Morokuma-Matsui, Maekawa, Muraoka, Nakajima, Umemoto, Sorai, Nakamura, Kuno, \& Saito}]{minamidani_development_2016}
Minamidani, T., Nishimura, A., Miyamoto, Y., {et~al.} 2016, 9914, 99141Z, \dodoi{10.1117/12.2232137}

\bibitem[{Mitchell {et~al.}(2021)Mitchell, Rowell, Celli, \& Einecke}]{mitchell_using_2021}
Mitchell, A. M.~W., Rowell, G.~P., Celli, S., \& Einecke, S. 2021, Monthly Notices of the Royal Astronomical Society, 503, 3522, \dodoi{10.1093/mnras/stab667}

\bibitem[{Mizuno {et~al.}(2022)Mizuno, Hayashi, Metzger, Moskalenko, Orlando, Strong, \& Yamamoto}]{mizuno_gas_2022}
Mizuno, T., Hayashi, K., Metzger, J., {et~al.} 2022, The Astrophysical Journal, 935, 97, \dodoi{10.3847/1538-4357/ac7de0}

\bibitem[{Morlino {et~al.}(2021)Morlino, Blasi, Peretti, \& Cristofari}]{morlino_particle_2021}
Morlino, G., Blasi, P., Peretti, E., \& Cristofari, P. 2021, Monthly Notices of the Royal Astronomical Society, 504, 6096, \dodoi{10.1093/mnras/stab690}

\bibitem[{Reid {et~al.}(2014)Reid, Menten, Brunthaler, Zheng, Dame, Xu, Wu, Zhang, Sanna, Sato, Hachisuka, Choi, Immer, Moscadelli, Rygl, \& Bartkiewicz}]{reid_trigonometric_2014}
Reid, M.~J., Menten, K.~M., Brunthaler, A., {et~al.} 2014, The Astrophysical Journal, 783, 130, \dodoi{10.1088/0004-637X/783/2/130}

\bibitem[{Reynolds {et~al.}(2023)Reynolds, Kara, Mushotzky, Ptak, Koss, Williams, Allen, Bauer, Bautz, Bodaghee, Burdge, Cappelluti, Cenko, Chartas, Chan, Corrales, Daylan, Falcone, Foord, Grant, Habouzit, Haggard, Herrmann, Hodges-Kluck, Kargaltsev, King, Kounkel, Lopez, Marchesi, McDonald, Meyer, Miller, Nynka, Okajima, Pacucci, Russell, Safi-Harb, Stassun, Falcão, Walker, Wilms, Yukita, \& Zhang}]{reynolds_overview_2023}
Reynolds, C.~S., Kara, E.~A., Mushotzky, R.~F., {et~al.} 2023, in {UV}, {X}-{Ray}, and {Gamma}-{Ray} {Space} {Instrumentation} for {Astronomy} {XXIII}, 49, \dodoi{10.1117/12.2677468}

\bibitem[{Safi-Harb {et~al.}(2023)Safi-Harb, Burdge, Bodaghee, An, Guest, Hare, Hebbar, Ho, Kargaltsev, Kirmizibayrak, Klingler, Nynka, Reynolds, Sasaki, Sridhar, Vasilopoulos, Woods, Yang, Heinke, Kong, Li, MacMaster, Mallick, Treyturik, Tsuji, Binder, Braun, Chang, Chatterjee, Ferrand, Holland-Ashford, Ng, Plotkin, Romani, \& Zhang}]{safi-harb_stellar_2023}
Safi-Harb, S., Burdge, K.~B., Bodaghee, A., {et~al.} 2023, From {Stellar} {Death} to {Cosmic} {Revelations}: {Zooming} in on {Compact} {Objects}, {Relativistic} {Outflows} and {Supernova} {Remnants} with {AXIS}, \dodoi{10.48550/arXiv.2311.07673}

\bibitem[{Sakemi {et~al.}(2023)Sakemi, Machida, Yamamoto, \& Tachihara}]{sakemi_molecular_2023}
Sakemi, H., Machida, M., Yamamoto, H., \& Tachihara, K. 2023, Publications of the Astronomical Society of Japan, 75, 338, \dodoi{10.1093/pasj/psad001}

\bibitem[{Sawada {et~al.}(2008)Sawada, Ikeda, Sunada, Kuno, Kamazaki, Morita, Kurono, Koura, Abe, Kawase, Maekawa, Horigome, \& Yanagisawa}]{sawada_--fly_2008}
Sawada, T., Ikeda, N., Sunada, K., {et~al.} 2008, Publications of the Astronomical Society of Japan, 60, 445, \dodoi{10.1093/pasj/60.3.445}

\bibitem[{Tanaka {et~al.}(2008)Tanaka, Uchiyama, Aharonian, Takahashi, Bamba, Hiraga, Kataoka, Kishishita, Kokubun, \& Mori}]{Tanaka2008}
Tanaka, T., Uchiyama, Y., Aharonian, F.~A., {et~al.} 2008, 685, 988, \dodoi{10.1086/591020}

\bibitem[{Taylor {et~al.}(2003)Taylor, Gibson, Peracaula, Martin, Landecker, Brunt, Dewdney, Dougherty, Gray, Higgs, Kerton, Knee, Kothes, Purton, Uyaniker, Wallace, Willis, \& Durand}]{taylor_canadian_2003}
Taylor, A.~R., Gibson, S.~J., Peracaula, M., {et~al.} 2003, The Astronomical Journal, 125, 3145, \dodoi{10.1086/375301}

\bibitem[{Tsuji {et~al.}(2021)Tsuji, Uchiyama, Khangulyan, \& Aharonian}]{tsuji_systematic_2021}
Tsuji, N., Uchiyama, Y., Khangulyan, D., \& Aharonian, F. 2021, The Astrophysical Journal, 907, 117, \dodoi{10.3847/1538-4357/abce65}

\bibitem[{Tsuji {et~al.}(2024)Tsuji, Tanaka, Safi-Harb, Aharonian, Casanova, Kothes, Moulin, Uchida, \& Uchiyama}]{tsuji_search_2024}
Tsuji, N., Tanaka, T., Safi-Harb, S., {et~al.} 2024, The Astrophysical Journal, 967, 138, \dodoi{10.3847/1538-4357/ad3fb1}

\bibitem[{Wenger {et~al.}(2018)Wenger, Balser, Anderson, \& Bania}]{wenger_kinematic_2018}
Wenger, T.~V., Balser, D.~S., Anderson, L.~D., \& Bania, T.~M. 2018, The Astrophysical Journal, 856, 52, \dodoi{10.3847/1538-4357/aaaec8}

\bibitem[{Wilson {et~al.}(2009)Wilson, Rohlfs, \& Hüttemeister}]{wilson_tools_2009}
Wilson, T.~L., Rohlfs, K., \& Hüttemeister, S. 2009, Tools of {Radio} {Astronomy}, \dodoi{10.1007/978-3-540-85122-6}

\bibitem[{Wouterloot \& Brand(1989)}]{wouterloot_iras_1989}
Wouterloot, J. G.~A., \& Brand, J. 1989, Astronomy and Astrophysics Supplement Series, 80, 149.
\newblock \url{https://ui.adsabs.harvard.edu/abs/1989A&AS...80..149W}

\end{thebibliography}
\bibliographystyle{aasjournal}



\end{document}